\documentclass{article} 
\usepackage{iclr2026_conference,times}

\usepackage{hyperref}
\usepackage{url}
\usepackage{algorithm}
\usepackage{algorithmic}
\usepackage{newfloat}
\usepackage{listings}

\usepackage{amsfonts}
\usepackage{amsmath}

\usepackage{booktabs}
\usepackage{multirow}
\usepackage{graphicx}
\usepackage{caption}
\usepackage{lscape}
\usepackage{rotating}
\usepackage{wrapfig}

\usepackage{xcolor}

\title{Step-Aware Residual-Guided Diffusion for EEG Spatial Super-Resolution}

\author{
  Hongjun Liu\textsuperscript{1}\thanks{Equal contribution.} \quad
  Leyu Zhou\textsuperscript{1}\footnotemark[1] \quad
  Zijianghao Yang\textsuperscript{1} \quad
  Chao Yao\textsuperscript{2}\thanks{Corresponding author.} \\
  \textsuperscript{1}School of Intelligence Science and Technology, University of Science and Technology Beijing \\
  \textsuperscript{2}School of Computer and Communication Engineering, University of Science and Technology Beijing \\
}

\iclrfinalcopy

\begin{document}

\maketitle

\begin{abstract}
For real-world {brain–computer interface (BCI)} applications, lightweight Electroencephalography (EEG) systems offer the best cost–deployment balance. However, such spatial sparsity of EEG limits spatial fidelity, hurting learning and introducing bias. EEG spatial super-resolution methods aim to recover high-density EEG signals from sparse measurements, yet is often hindered by distribution shift and signal distortion and thus reducing fidelity and usability for EEG analysis and visualization. 
To overcome these challenges, we introduce SRGDiff, a step-aware residual-guided diffusion model that formulates EEG spatial super-resolution as dynamic conditional generation.
Our key idea is to learn a dynamic residual condition from the low-density input that predicts the step-wise temporal and spatial details to add and uses the evolving cue to steer the denoising process toward high density reconstructions.
At each denoising step, the proposed residual condition is additively fused with the previous denoiser feature maps, then a step-dependent affine modulation scales and shifts the activation to produce the current features. 
This iterative procedure dynamically extracts step-wise temporal rhythms and spatial-topographic cues to steer high-density recovery and maintain a fidelity–consistency balance.
We adopt a comprehensive evaluation protocol spanning signal-, feature-, and downstream-level metrics across SEED, SEED\mbox{-}IV, and Localize\mbox{-}MI and multiple upsampling scales. 
{SRGDiff consistently achieves higher SNR than the baseline ESTformer and STAD among Localize-MI, SEED and SEED-IV datasets, with up to roughly $75\%$ relative SNR improvement in the most challenging $8\times$ setting}.
Moreover, topographic visualizations comparison and substantial EEG-FID gains jointly indicate that our SR EEG mitigates the spatial–spectral shift between low- and high-density recordings. Our code is available at https://github.com/DhrLhj/ICLR2026SRGDiff.

\end{abstract}

\section{Introduction}
\begin{wrapfigure}{l}{0.6\linewidth} 
  \vspace{-0.6\baselineskip}         
  \centering
  \includegraphics[width=\linewidth]{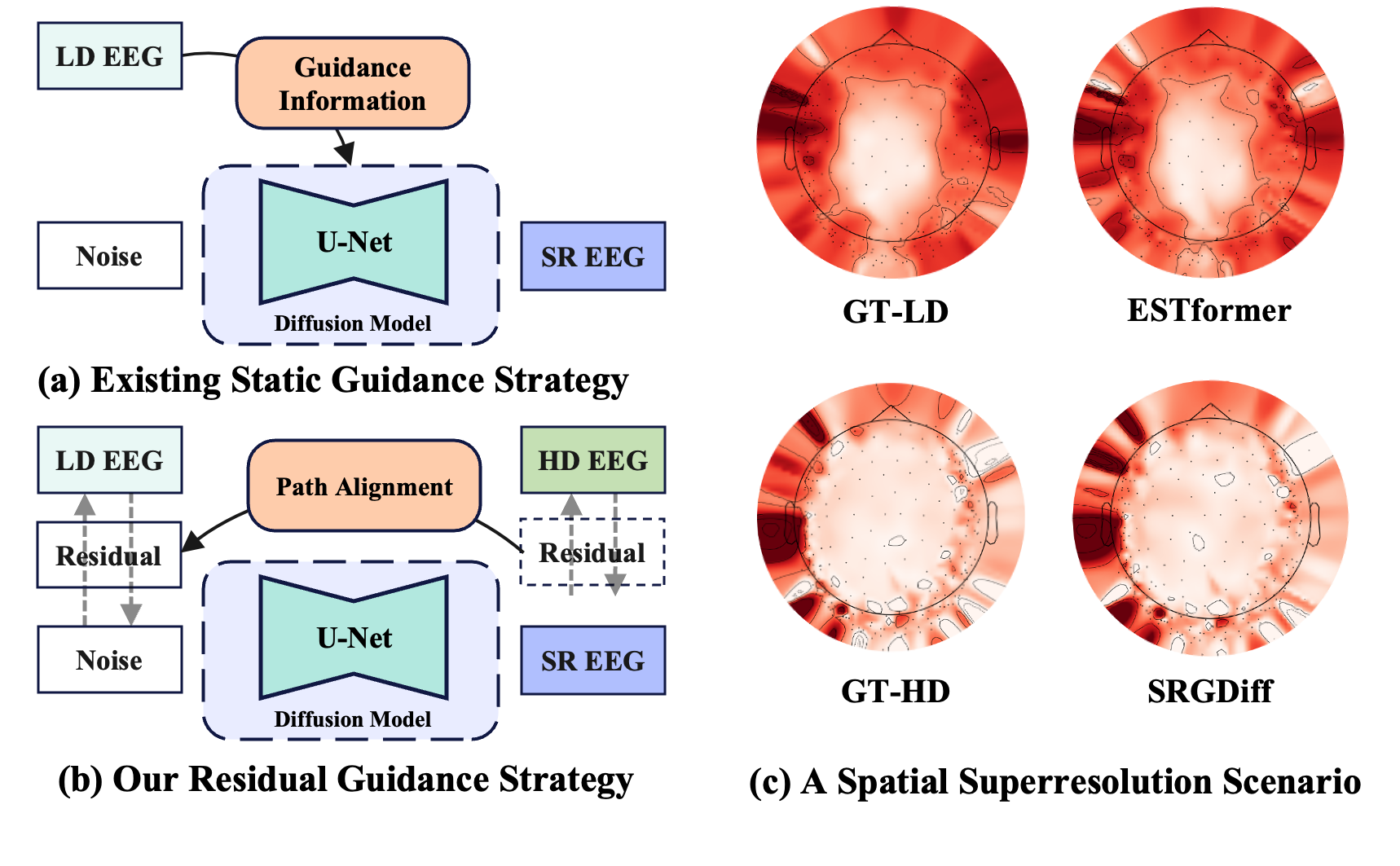}
  \caption{(a) Existing static guidance strategy vs. (b) our residual guidance strategy for EEG super-resolution, and (c) corresponding topographical maps of LD input, ESTformer output, GT HD EEG, and SRGDiff reconstruction.}
  \label{fig:illustration}
\end{wrapfigure}

Electroencephalography (EEG) is a noninvasive technique for monitoring the brain’s electrical activity, with widespread applications in neuroscience and clinical practice—ranging from brain–computer interfaces and epilepsy diagnosis to emotion recognition {\citep{jiangneurolm}}. However, EEG’s spatial resolution is inherently constrained by the number of scalp electrodes and the volume‐conduction effect {\citep{li2025estformer}}. High‐density (HD) systems with hundreds of channels can mitigate these issues but are costly, cumbersome to deploy, and uncomfortable for extended wear, whereas low‐density (LD) setups (e.g., 8 or 16 electrodes) are far more practical yet suffer from severe under‐sampling bias {\citep{wang2025generative}}. Indeed, as illustrated in Figure \ref{fig:illustration}(c), the inter‐channel activation patterns of 256‐channel HD EEG diverge dramatically from those of 16‐channel LD EEG, highlighting the strong bias in sparse recordings. EEG spatial super-resolution (SR) has therefore garnered growing attention, with methods that reconstruct high-density EEG from sparse recordings increasingly explored and applied.

Traditionally, EEG spatial super‐resolution has relied on direct feature‐mapping techniques that learn an end-to-end mapping from low-density to high-density representations. These methods fall into two main categories: one employs convolutional neural networks or Transformers to upsample LD feature maps into HD ones {\citep{tang2022deep}}, and the other leverages generative adversarial networks-based architectures that synthesize SR EEG signals conditioned on LD inputs {\citep{wang2024eeg}}. However, by treating the mapping as a static projection, these approaches often oversimplify the complex, nonlinear inter-electrode dependencies or demand vast amounts of training data and compute, resulting in overly smooth, detail-poor reconstructions that fail to capture true spatial consistency. Figure~\ref{fig:illustration}(c) indicates that such feature-mapping methods merely extend LD information, rather than recovering authentic HD channel relationships.

Recently, diffusion models have been widely applied to time-series generation and missing-data imputation {\citep{huang2025timedp, yuan2024diffusionts, li2025population}}. In this context, EEG spatial super-resolution can be cast as conditional generation, where LD observations guide the recovery of HD signals.
Within this line of work, researchers mainly focused on conditioning strategies through concatenating low-density features with the noise input {\citep{vetter2024generating}} or using cross-attention between modalities {\citep{wang2025generative}} as shown in Figure~\ref{fig:illustration}(a).
While effective in practice, these approaches remain susceptible to a consistency–fidelity trade-off. Interpolation-oriented SR tends to cause distribution shift, making reconstructions adhere too closely to the LD observation and deviate from the HD ground truth. Conversely, generation-oriented SR often introduces distortion, producing HD-like content that fails to remain consistent with the LD input.

To tackle these challenges, we introduce Step-aware Residual-Guided Diffusion (SRGDiff) for EEG Spatial Super-Resolution, which reframes super-resolution as a dynamic conditional generation task. The core idea is to estimate the forward-noising residual from low-density channels, and use it as a per-step corrective direction in the reverse process.
Technically, SRGDiff first encodes the low-density EEG with a pre-trained VAE encoder to obtain a compact latent and multi-scale features, and applies forward diffusion to the high-density latent. At each reverse step, a lightweight residual head predicts a path residual from the low-density features and uses it as a directional correction that is additively fused with the previous denoising features to form an incremental feature. The feature is then weighted with a step-dependent affine modulation estimated from the low-density features and the timestep embedding, yielding the current denoised features. This loop repeats over timesteps, coupling the low-density forward-noising and high-density reverse-denoising trajectories and progressively steering samples toward the high-density manifold. Our main contributions in this work can be summarized as follows:
\begin{itemize}
    \item We recast EEG spatial super-resolution as \textbf{dynamic conditional generation}, coupling the LD forward–noising trajectory with the HD reverse–denoising trajectory to balance consistency with the LD observation and fidelity to the HD target.
    \item We propose a \textbf{dynamic residual guidance} paradigm: the path residual estimated from LD inputs serves as a per-step directional correction and is fused additively for incremental updates, {yielding a stable, step-aware sampling scheme that remains effective across datasets and a wide range of SR factors.}
    \item We establish a \textbf{three-level evaluation protocol} across three datasets, covering signal-level (temporal consistency, spectral fidelity, spatial topology), feature-level (representation quality), and downstream-level (classification accuracy), which provides a comprehensive assessment beyond pointwise error.
\end{itemize}

\section{Related Work}

\paragraph{Diffusion Models for Missing Data Imputation.} Diffusion‐based models have emerged as a powerful framework for time‐series imputation, leveraging denoising diffusion processes to reconstruct missing values. Diffusion-TS {\citep{yuan2024diffusionts}} demonstrates interpretable conditional generation across diverse time‐series without domain‐specific priors. More recently, RDPI {\citep{liu2025rdpi}} further enhances precision and efficiency by first generating coarse estimates of missing values through deterministic interpolation, then conditioning a diffusion model on both observed data and these estimates to iteratively refine residual errors. 
SaSDim {\citep{ijcai2024p283}} introduces self‑adaptive noise scaling to preserve spatial dependencies within sensor networks, while SADI {\citep{AAAI2025sadi}} integrate self‑attention mechanisms to handle partial data missing.

\paragraph{EEG Spatial Super-Resolution.} Early attempts at EEG spatial super‐resolution adapted image‐based frameworks to reconstruct dense electrode maps from sparse recordings. EEGSR‐GAN {\citep{8333379}} first applied adversarial training to hallucinate missing channels. ESTformer {\citep{li2025estformer}} then introduced spatiotemporal transformers to model long‐range dependencies across electrodes and successfully capture global patterns.
More recent diffusion‐based and attention‐driven approaches have sought to address these limitations. DDPM‐EEG {\citep{vetter2024generating}} leverages denoising diffusion probabilistic models to iteratively refine spatial patterns. STAD {\citep{wang2025generative}} tackles this by decomposing spatial–temporal interactions into spatial-temporal attention streams. This diffusion-based generative paradigm improves diversity and spectral fidelity compared to GANs, yet their static condition can still lead to distribution drifts and distortion. 

\paragraph{Residual Diffusion in Related Domains.}
Several recent works incorporate residual signals into diffusion models. 
\cite{ou2024prior} synthesize PET from MRI by learning a modality residual under prior information, i.e., a static cross-modal gap that conditions generation. \cite{zhu2024temporal} reconstruct event-driven video by predicting temporal residuals with inter-frame differences as the generation target to recover dynamics. \cite{mao2025prior} address medical segmentation by learning a residual-to-prior that corrects a coarse segmentation, improving calibration and efficiency. These designs either treat the residual as a fixed target/offset or inject it once as a global prior, with weak coupling to the step-by-step reverse dynamics. In contrast, our method targets EEG spatial super-resolution and introduces a dynamic, step-aware residual direction that is re-estimated at every reverse step from the LD observation, and the timestep embedding.

\begin{figure}[]
  \centering
  \includegraphics[width=\columnwidth]{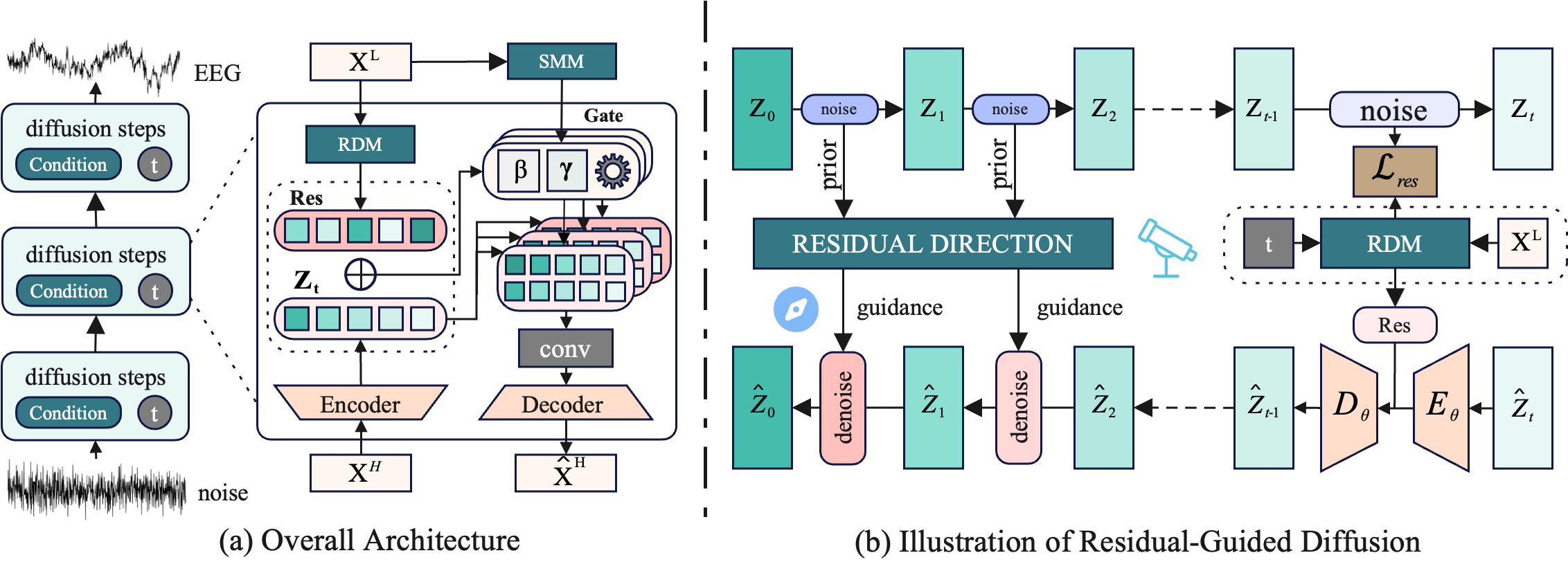}
  \caption{\textbf{SRGDiff overview.} 
(a) \textbf{Overall architecture:} Low-density EEG $X^{L}$ conditions the latent reverse process. 
RDM predicts a residual direction from $X^{L}$ and current decoder features.
SMM provides step-aware affine parameters to fuse the residual and modulate activations.  
(b) \textbf{Residual learning:} At each step, the predicted residual guides denoising, and the residual derived from the forward noising process provides supervision via a residual loss.
}
  \label{fig:overview}
\end{figure}  

\section{Preliminaries and Dynamic Conditional Formulation}

\noindent\textbf{Data and Latent Space.}
Let $X^L\!\in\!\mathbb{R}^{C_L\times Length}$ and $X^H\!\in\!\mathbb{R}^{C_H\times Length}$ denote low- and high-density EEG with $C_H\!>\!C_L$.
A pre-trained VAE encoder $E$ maps signals to a latent space $ z = E(X^H)$ and a feature extractor $F$ provides LD features as condition $c = F(X^L)$ to condition generation. In practice, we reuse the VAE encoder as the feature extractor to obtain condition $c$.

\medskip
\noindent\textbf{Forward Diffusion on the HD Latent.}
We corrupt the HD latent with a standard Gaussian forward process
\begin{equation}
q(z_t \mid z_{z-1}) = \mathcal{N}\!\Big(\sqrt{\alpha_t}\,z_{t-1},\, (1-\alpha_t) I\Big).\qquad t=1,\dots,T
\end{equation}

\medskip
\noindent\textbf{Dynamic Conditional Reverse Process.} 
In the reverse generation stage, sampling begins from an isotropic Gaussian noise initialization $\hat{z}_T \sim \mathcal{N}(0,I)$, and the model iteratively predicts $\hat{z}_{t-1}$ from $\hat{z}_t$ until recovering the final latent representation $\hat{z}_0$. Unlike conventional diffusion models that rely solely on a base denoiser, SRGDiff explicitly conditions each reverse step on low-density EEG observations, thereby coupling the LD forward-noising trajectory with the HD reverse-denoising trajectory. The reverse denoiser is defined as:
\begin{equation}
p_\theta(\hat{z}_{t-1} \mid \hat{z}_t, c) 
= \mathcal{N}\!\Big( 
\underbrace{\mu_\theta(\hat{z}_t, c)}_{\text{base denoiser}} 
+ \underbrace{(\gamma_t, \beta_t, r_\phi(c,t))}_{\text{dynamic conditional update}},\,
\beta_t I \Big).
\end{equation}

Here, the base denoiser $\mu_\theta(\hat{z}_t, c)$ is implemented as a U\mbox{-}Net that estimates the noise component. To further enhance temporal fidelity and spatial coherence, we augment the base denoiser with two lightweight modules that inject step-wise conditional guidance from LD features:

\begin{itemize}
    \item \textit{Residual Direction Module (RDM).} 
At each timestep, RDM predicts a path residual $r_\phi(c,t)$ from the LD features and applies it as a directional correction:
\begin{equation}
\hat{z}^{RDM}_{t-1} = \hat{z}_t + r_\phi(c,t).
\end{equation}
\item \textit{Step-Aware Modulation Module (SMM).} 
SMM calibrates the residual update with timestep-aware affine modulation. Specifically, it predicts a scale $\gamma_t$ and bias $\beta_t$ from LD features and the timestep embedding, and applies them to the residual-corrected state:
\begin{equation}
\hat{z}^{SMM}_{t-1} = \gamma_t \odot \hat{z}^{RDM}_{t-1} + \beta_t.
\end{equation}
\end{itemize}

Together, the pair $(r_\phi, \gamma_t, \beta_t)$ realizes dynamic conditional generation: at every denoising step, LD features provide both a directional residual and a step-dependent modulation strength, yielding a stable and temporally consistent correction of the reverse diffusion process.

\section{Proposed Method}

This section presents the step-aware residual-guided diffusion framework, and outlines its architecture and core components. SRGDiff reframes SR as dynamic conditional generation that couples the LD forward-noising trajectory with the HD reverse-denoising trajectory, using an LD-estimated residual as a per-step corrective direction and a step-aware calibration to modulate its strength. The framework comprises four parts: the latent diffusion model backbone, Residual Direction Module (RDM) for additive directional updates, Step-Aware Modulation Module (SMM) for step-dependent modulation, and the overall training strategy. An overview is provided in Figure \ref{fig:overview}, and the following subsections describe each component in detail.

\subsection{Latent Diffusion Model Backbone}

Our backbone consists of a VAE that builds the latent space and a denoising U\mbox{-}Net that performs diffusion in that space. 
The VAE follows the EEG autoencoding setup of \cite{aristimunha2023synthetic}. We train an encoder–decoder $(E,D)$ on HD EEG $X^H$ to obtain $z=E(X^H)$ and $\widehat X^H=D(\widehat z)$, and optimize a reconstruction–regularization objective
\begin{equation}
\mathcal{L}_{\text{VAE}}
= \|\widehat X^H - X^H\|_2^2
+ \lambda_{\text{spec}}\big\|\mathrm{STFT}(\widehat X^H)-\mathrm{STFT}(X^H)\big\|_1
+ \lambda_{KL}\,\mathrm{KL}\!\left(q_E(z\!\mid\!X^H)\,\|\,\mathcal{N}(0,I)\right),
\end{equation}
{where $\mathrm{STFT}(\cdot)$ denotes the short\mbox{-}time Fourier transform applied along the temporal dimension of each EEG channel to encourage spectral fidelity.}

Empirically, we set the spectral weight to $0.1$ and the KL weight to $10^{-4}$. After convergence, $(E,D)$ are frozen.
On top of this latent space, we adopt a latent\mbox{-}diffusion model in the style of \cite{rombach2022high}.

\subsection{Residual Direction Module}

In the context of EEG spatial SR, most diffusion approaches condition the U\mbox{-}Net via feature concatenation or cross-attention. We instead turn EEG spatial SR into finer and step-aware conditioning by learning residual direction from low-density recordings. We use the VAE encoder to extract multi-scale condition $c$ and an RDM head $R_{\phi}$ takes $(c,\,\tau(t))$ to predict a residual $Res_{t}$ in the encoder feature space, which acts as a per-step directional correction to the reverse process. Concretely, we first sample a timestep $t$, encode the HD EEG to obtain the latent $z_0=E(X^H)$, and draw
\begin{equation}
z_t \;=\; \sqrt{\bar\alpha_t}\,z_0 \;+\; \sqrt{1-\bar\alpha_t}\,\epsilon,\qquad \epsilon\sim\mathcal N(0,I).
\end{equation}

In the forward process, we obtain at each step the noise-corrupted latent of the HD EEG and use this sequence of step-dependent features as supervision targets for the residual. The residual labels span $t=0,\dots,T$ and are defined as $ \delta z_t \;:=\; z_0 \;-\; z_t, $. 

In the reverse process, we estimate $\delta z_t$ from the low\mbox{-}density EEG to supply the step\mbox{-}wise temporal and spatial details required for denoising. 
We introduce a lightweight convolutional predictor $R_{\phi}$ that takes the timestep embedding $\tau(t)$ and LD features $c=F(X^{L})$ as input and outputs the residual feature $Res_{t}$, and trained by
\begin{equation}
Res_{t} \;=\; R_{\phi}\!\big(\tau(t),\,c\big), \quad
\mathcal{L}_{\mathrm{res}}
\;=\; \sum_{t=0}^{T} \big\|\,Res_{t} - \delta z_t\big\|_{2}^{2}.
\end{equation}

Finally, the predicted residual feature is then added to $\hat z_{t}$ as an incremental update:
\begin{equation}
\hat z_{t}^{RDM} \;=\; LayerNorm(\hat z_{t}) \;+\; Res_t.
\end{equation}

\subsection{Step-Aware Modulation Module}

After obtaining $\hat z_{t}^{\mathrm{RDM}}$, we further modulate the current step to control the extent to which the residual condition influences denoising. 
To enforce temporal fidelity, SMM explicitly weighted the current diffusion timestep with a step-dependent affine modulation estimated from the low-density features and the current timestep embedding. 

Specifically, SMM first encodes the low‐density EEG through a lightweight 1D convolutional network $E_{SMM}$ to produce a feature map \(h_t\). To enable the conditioning to recognize the current diffusion step, SMM maps each sampled timestep $t$ into a sinusoidal time embedding $e_t$. A learnable weight \(\sigma_t\) that decays linearly with \(t\) balances these two streams, yielding a fused feature:
\begin{equation}\widetilde h_t = \sigma_t h_t + (1-\sigma_t)e_t = \sigma_t E_{SMM}(c) + (1-\sigma_t)e_t.\end{equation} 

For spatial coherence, we adopt an affine calibration mechanism. The fused feature \(\widetilde h_t\) is passed through two MLPs $MLP_{\gamma}$ and $MLP_{\beta}$ to predict channel‐wise scale \(\gamma^c_t\) and bias \(\beta^c_t\):
\begin{equation}
\hat z_t^{SMM} = \gamma_t \,\odot\, \hat z_t^{RDM}  \;+\;\beta^c_t=MLP_{\gamma}(\widetilde h_t, t) \,\odot\, \hat z_t^{RDM} \;+\; MLP_{\beta}(\widetilde h_t,t).
\end{equation}
Finally, SRGDiff feeds the updated latent $\hat z_t$ into the U\mbox{-}Net decoder to obtain the next denoised state $\hat z_{t-1}$.

\subsection{Training Strategy}

To stabilize optimization and decouple latent representation learning from conditional diffusion modeling, we adopt a two-stage training strategy.

\textbf{Stage 1: VAE Pre-training.} 
We first train the VAE encoder-decoder pair on high-density EEG data to obtain a stable and structured latent space. After convergence, the VAE parameters are frozen to provide fixed latent representations for subsequent diffusion modeling.

\textbf{Stage 2: Residual-Guided Latent Diffusion.}
On the frozen latent space, the final training objective is a weighted combination of the three terms:
\begin{equation}
\mathcal{L}_{\text{Stage 2}}
= \mathbb{E}_{z_0,\epsilon,t}\big[ \|\epsilon - \epsilon_\theta(z_t, t, c)\|_2^2 \big]
+ \lambda_{res}\sum_{t=1}^T \| R_\varphi(c, t) - (z_0 - z_t) \|_2^2
+ \lambda_{SMM}(\|\gamma_t - 1\|_2^2 + \|\beta_t\|_2^2).
\end{equation}
Empirically, we set the residual weight to $1$ and the SMM weight to $10^{-2}$. The term $\lambda_{\text{SMM}}\big(\|\gamma_t - 1\|_2^2 + \|\beta_t\|_2^2\big)$ serves as a regularization component to prevent excessively large values of $\gamma_t$ and $\beta_t$, thereby stabilizing the training dynamics.

\section{Experiments}

\subsection{Downstream Datasets}

In this study, we employ three publicly available EEG datasets. The \textbf{SEED dataset} {\citep{zheng2015investigating}} uses $15$ film clips of approximately four minutes each as emotional stimuli to induce stable and continuous emotional responses in three categories: positive, neutral, and negative. Data were acquired via $62$ channels at $1000$ Hz, downsampled to $200$ Hz, band-pass filtered ($0$–$75$ Hz) and segments with faulty sensors removed. 
\textbf{SEED-IV} {\citep{8283814}} extends SEED by using the same $15$ subjects and $62$-channel setup ($1000$ Hz) but adds music and image stimuli to evoke happiness, sadness, fear and neutral states; preprocessing mirrors that of SEED. 
\textbf{Localize-MI} {\citep{mikulan2020simultaneous}} contains 61 presurgical sessions from seven drug-resistant epilepsy patients, where $256$-channel scalp EEG was recorded at $8000$ Hz during $0.1$–$5$ mA intracerebral single-pulse stimulation; preprocessing includes a $0.1$ Hz high-pass filter, notch filter, bad-channel/trial removal and alignment of trials to the -$300$ ms to +$50$ ms stimulus-artifact window.

\subsection{Experiment Setup}

\paragraph{Data Preprocessing.} The experimental setup for the EEG super-resolution task follows the ESTformer and STAD frameworks. The preprocessed EEG signals were segmented into fixed-length windows: continuous, non-overlapping $4$-second windows for SEED and SEED-IV datasets, and $260$ ms windows (from $250$ ms before stimulation to $10$ ms after) for Localize-MI. In SEED and SEED-IV, we designed different super-resolution scale factors ($2\times$, $4\times$ and $8\times$) to evaluate reconstruction performance. The selection of visible channels and the super-resolution scaling factors follow the configurations used in ESTformer. For Localize-MI, due to the high channel density, we applied more extensive scale factors ($2\times$, $4\times$, $8\times$, $16\times$). 

\paragraph{Training \& Environment Settings.} For each dataset, we split the data into train/test with an $80\%/20\%$ ratio and reserve $10\%$ of the training portion as a validation set. Stage I is trained only on the HD signals from the training split. Stage II is trained on paired LD/HD samples constructed from the same training split by masking HD channels according to the target SR scale; the validation set is used for early stopping and hyperparameter selection. The held-out test split is used once for final reporting, with no fine-tuning.

\paragraph{Baselines.}
We compare SRGDiff with strong EEG SR and time-series imputation baselines: \textbf{ESTformer} {\citep{li2025estformer}} and \textbf{STAD} {\citep{wang2025generative}} (transformer-/diffusion-based EEG SR), \textbf{DDPMEEG} {\citep{vetter2024generating}} (diffusion for ECoG SR), \textbf{SaSDim} {\citep{ijcai2024p283}} and \textbf{SADI} {\citep{AAAI2025sadi}} (advanced missing data imputation), and the two-stage residual method \textbf{RDPI} {\citep{liu2025rdpi}}. 
We use authors’ official implementations when available and otherwise provide carefully verified reimplementations, applying their recommended hyperparameters and unifying training epochs and sampling steps across methods.

\begin{table*}[!t]
\small
  \centering
  {
  \begin{tabular}{llccccccc}
    \toprule
    \textbf{Model}   & \textbf{Ref}        & \textbf{Metric}
      & \textbf{2}       & \textbf{4}
      & \textbf{8}       & \textbf{16} \\
    \midrule

    \multirow{3}{*}{SaSDim}
      & \multirow{3}{*}{IJCAI 2024} & NMSE
        & 0.2675\scriptsize{$\pm0.003$}
        & 0.3427\scriptsize{$\pm0.001$}
        & 0.4174\scriptsize{$\pm0.004$}
        & 0.4613\scriptsize{$\pm0.003$} \\
      &                             & PCC
        & 0.8194\scriptsize{$\pm0.002$}
        & 0.7246\scriptsize{$\pm0.007$}
        & 0.6926\scriptsize{$\pm0.003$}
        & 0.6476\scriptsize{$\pm0.002$} \\
      &                             & SNR
        & 5.7443\scriptsize{$\pm0.007$}
        & 4.3796\scriptsize{$\pm0.003$}
        & 3.5549\scriptsize{$\pm0.009$}
        & 2.7678\scriptsize{$\pm0.005$} \\

    \midrule

    \multirow{3}{*}{SADI}
      & \multirow{3}{*}{AAAI 2025} & NMSE
        & 0.2637\scriptsize{$\pm0.003$}
        & 0.3442\scriptsize{$\pm0.001$}
        & 0.4164\scriptsize{$\pm0.004$}
        & 0.4566\scriptsize{$\pm0.003$} \\
      &                            & PCC
        & 0.8243\scriptsize{$\pm0.002$}
        & 0.7391\scriptsize{$\pm0.007$}
        & 0.6944\scriptsize{$\pm0.003$}
        & 0.6554\scriptsize{$\pm0.002$} \\
      &                            & SNR
        & 5.7511\scriptsize{$\pm0.007$}
        & 4.3724\scriptsize{$\pm0.003$}
        & 3.5498\scriptsize{$\pm0.008$}
        & 2.8942\scriptsize{$\pm0.009$} \\

    \midrule
    
    \multirow{3}{*}{RDPI}
      & \multirow{3}{*}{AAAI 2025} & NMSE
        & 0.2561\scriptsize{$\pm0.003$}
        & 0.3562\scriptsize{$\pm0.001$}
        & 0.4076\scriptsize{$\pm0.004$}
        & 0.4531\scriptsize{$\pm0.003$} \\
      &                            & PCC
        & 0.8246\scriptsize{$\pm0.002$}
        & 0.7396\scriptsize{$\pm0.007$}
        & 0.7062\scriptsize{$\pm0.003$}
        & 0.6549\scriptsize{$\pm0.002$} \\
      &                            & SNR
        & 5.7311\scriptsize{$\pm0.007$}
        & 4.3966\scriptsize{$\pm0.003$}
        & 3.5643\scriptsize{$\pm0.009$}
        & 2.7731\scriptsize{$\pm0.007$} \\
    \midrule
        \multirow{3}{*}{DDPMEEG}
      & \multirow{3}{*}{Patterns 2024} & NMSE
        & 0.2046\scriptsize{$\pm0.003$}
        & 0.3108\scriptsize{$\pm0.001$}
        & \underline{0.3554}\scriptsize{$\pm0.004$}
        & \underline{0.4076}\scriptsize{$\pm0.002$} \\
      &                               & PCC
        & 0.8516\scriptsize{$\pm0.002$}
        & 0.8163\scriptsize{$\pm0.007$}
        & \underline{0.7306}\scriptsize{$\pm0.003$}
        & \underline{0.6739}\scriptsize{$\pm0.002$} \\
      &                               & SNR 
        & 6.2151\scriptsize{$\pm0.008$}
        & 5.5126\scriptsize{$\pm0.003$}
        & \underline{3.9891}\scriptsize{$\pm0.009$}
        & \underline{3.2715}\scriptsize{$\pm0.005$} \\

    \midrule
    
    \multirow{3}{*}{ESTformer}
      & \multirow{3}{*}{KBS 2025} & NMSE
        & 0.2721\scriptsize{$\pm0.003$}
        & 0.3578\scriptsize{$\pm0.001$}
        & 0.4466\scriptsize{$\pm0.004$}
        & 0.4837\scriptsize{$\pm0.002$} \\
      &                            & PCC
        & 0.8061\scriptsize{$\pm0.002$}
        & 0.7205\scriptsize{$\pm0.007$}
        & 0.6867\scriptsize{$\pm0.003$}
        & 0.6319\scriptsize{$\pm0.002$} \\
      &                            & SNR
        & 5.5403\scriptsize{$\pm0.008$}
        & 3.8671\scriptsize{$\pm0.003$}
        & 3.3023\scriptsize{$\pm0.007$}
        & 2.5671\scriptsize{$\pm0.004$} \\
    
    \midrule

    \multirow{3}{*}{STAD}
      & \multirow{3}{*}{TCE 2025} & NMSE
        & \underline{0.1902}\scriptsize{$\pm0.003$}
        & \underline{0.3067}\scriptsize{$\pm0.001$}
        & 0.3649\scriptsize{$\pm0.004$}
        & 0.4106\scriptsize{$\pm0.003$} \\
      &                           & PCC
        & \underline{0.8635}\scriptsize{$\pm0.002$}
        & \underline{0.8194}\scriptsize{$\pm0.007$}
        & 0.7216\scriptsize{$\pm0.003$}
        & 0.6694\scriptsize{$\pm0.002$} \\
      &                           & SNR
        & \underline{7.2591}\scriptsize{$\pm0.008$}
        & \underline{5.5234}\scriptsize{$\pm0.003$}
        & 3.8715\scriptsize{$\pm0.009$}
        & 3.2642\scriptsize{$\pm0.005$} \\

    \midrule
    
    \multirow{3}{*}{SRGDiff}
      & \multirow{3}{*}{OURS}          & NMSE
        & \textbf{0.1449}\scriptsize{$\pm0.003$}
        & \textbf{0.2384}\scriptsize{$\pm0.001$}
        & \textbf{0.2957}\scriptsize{$\pm0.004$}
        & \textbf{0.3457}\scriptsize{$\pm0.002$} \\
      &                            & PCC
        & \textbf{0.9213}\scriptsize{$\pm0.002$}
        & \textbf{0.8854}\scriptsize{$\pm0.007$}
        & \textbf{0.8323}\scriptsize{$\pm0.003$}
        & \textbf{0.7322}\scriptsize{$\pm0.002$} \\
      &                            & SNR
        & \textbf{8.3755}\scriptsize{$\pm0.008$}
        & \textbf{6.3617}\scriptsize{$\pm0.003$}
        & \textbf{5.2249}\scriptsize{$\pm0.009$}
        & \textbf{4.0197}\scriptsize{$\pm0.006$} \\

    \bottomrule
  \end{tabular}
  }
  \caption{Performance of all methods on Localize-MI across different channel settings.}
  \label{tab:MISR}
\end{table*}

\subsection{Evaluation Protocol. } 

We assess SR quality at three complementary levels to balance faithfulness to the ground truth, preservation of neurophysiological structure, and practical utility. 

\paragraph{Signal level} (does the waveform match?): We follow \textbf{ESTformer} and report {normalized mean squared error (NMSE), Pearson correlation coefficient (PCC), and reconstruction signal-to-noise ratio (SNR) with respect to the HD reference}, plus topology maps for qualitative inspection (formal definitions in the Appendix). 

\paragraph{Feature level} (does the representation distribution match?): We adopt \textbf{EEG-FID} following \cite{lai2025diffusets}, using a frozen EEGNet trained per dataset on its training split; the embedding dimension is 256 for SEED/SEED-IV and 512 for Localize-MI. {In addition, we report a \textbf{frequency-domain MAE}: we first compute channel-wise STFTs of the reconstructed and reference HD EEG, form their power spectra, and then compute the normalized mean squared error between these spectra averaged over channels and frequency bins, so as to capture spectral distortions that are not reflected by time-domain NMSE alone.}

\paragraph{Downstream level} (is it useful?): We evaluate SEED/SEED-IV subject-dependent emotion recognition without cross-validation and binary epileptic classification on Localize-MI, both reporting accuracy. 
All results are summarized as mean$\pm$std over subjects; implementation details and metric formulas are provided in the Appendix.

\subsection{Main Results}

We report signal-level reconstruction quality in Tables~\ref{tab:MISR} and \ref{tab:SeedSR}. 
For the most demanding $16\times$ upsampling on Localize\mbox{-}MI, SRGDiff attains an NMSE of $0.3457$, over $15\%$ lower than DDPMEEG’s $0.4076$, indicating that dynamic conditioning effectively guides the diffusion model to generate super\mbox{-}resolved signals that closely approximate the true high\mbox{-}density data. Its PCC improves from $0.6739$ to $0.7322$, and its SNR increases from $3.27$\,dB to $4.02$\,dB (over $22\%$), demonstrating that under challenging settings the dynamic conditioning still learns the HD trend while maintaining a favorable signal\mbox{-}to\mbox{-}noise ratio. 
On SEED with  high temporal variability and frequent outliers, SRGDiff reduces the $2\times$ NMSE from ESTformer’s $0.3288$ to $0.1632$ (a reduction of more than $50\%$) and raises PCC from $0.8368$ to $0.9102$. A similar pattern is observed on SEED\mbox{-}IV, where NMSE drops to $0.1663$ versus $0.3448$ and PCC increases to $0.9113$ versus $0.8106$, indicating that despite lower SNR, the dynamic conditioning exhibits strong generalization.


\begin{table*}[!t]
\centering
\small
\resizebox{\linewidth}{!}{
\begin{tabular}{llcccccc}
\toprule
\multirow{2}{*}{\textbf{Model}} & \multirow{2}{*}{\textbf{Metric}} & \multicolumn{3}{c}{\textbf{SEED (62)}} & \multicolumn{3}{c}{\textbf{SEED-IV (62)}} \\
\cmidrule(lr){3-5} \cmidrule(lr){6-8}
& & \textbf{2} & \textbf{4} & \textbf{8} & \textbf{2} & \textbf{4} & \textbf{8} \\
\midrule
\multirow{3}{*}{SaSDim}
& NMSE & 0.4399{\scriptsize{$\pm$0.004}} & 0.6234{\scriptsize{$\pm$0.002}} & 0.7767{\scriptsize{$\pm$0.007}} & 0.3633{\scriptsize{$\pm$0.004}} & 0.5543{\scriptsize{$\pm$0.002}} & 0.7122{\scriptsize{$\pm$0.005}} \\
& PCC  & 0.7341{\scriptsize{$\pm$0.002}} & 0.5649{\scriptsize{$\pm$0.001}} & 0.4349{\scriptsize{$\pm$0.004}} & 0.7249{\scriptsize{$\pm$0.002}} & 0.6211{\scriptsize{$\pm$0.009}} & 0.5009{\scriptsize{$\pm$0.003}} \\
& SNR  & 4.1154{\scriptsize{$\pm$0.096}} & 2.2940{\scriptsize{$\pm$0.046}} & 1.1349{\scriptsize{$\pm$0.127}} & 4.5940{\scriptsize{$\pm$0.009}} & 2.6004{\scriptsize{$\pm$0.004}} & 1.6211{\scriptsize{$\pm$0.111}} \\
\midrule

\multirow{3}{*}{SADI}
& NMSE & 0.4439{\scriptsize{$\pm$0.004}} & 0.6049{\scriptsize{$\pm$0.002}} & 0.8106{\scriptsize{$\pm$0.007}} & 0.3557{\scriptsize{$\pm$0.004}} & 0.5349{\scriptsize{$\pm$0.002}} & 0.6844{\scriptsize{$\pm$0.005}} \\
& PCC  & 0.7234{\scriptsize{$\pm$0.002}} & 0.5819{\scriptsize{$\pm$0.001}} & 0.4064{\scriptsize{$\pm$0.004}} & 0.7624{\scriptsize{$\pm$0.002}} & 0.6293{\scriptsize{$\pm$0.009}} & 0.5243{\scriptsize{$\pm$0.003}} \\
& SNR  & 4.2419{\scriptsize{$\pm$0.097}} & 2.5160{\scriptsize{$\pm$0.046}} & 1.0137{\scriptsize{$\pm$0.127}} & 4.7093{\scriptsize{$\pm$0.009}} & 2.6044{\scriptsize{$\pm$0.004}} & 1.6610{\scriptsize{$\pm$0.112}} \\
\midrule

\multirow{3}{*}{RDPI}
& NMSE & 0.4064{\scriptsize{$\pm$0.004}} & 0.6134{\scriptsize{$\pm$0.002}} & 0.7916{\scriptsize{$\pm$0.007}} & 0.3491{\scriptsize{$\pm$0.004}} & 0.5416{\scriptsize{$\pm$0.002}} & 0.6915{\scriptsize{$\pm$0.005}} \\
& PCC  & 0.7416{\scriptsize{$\pm$0.002}} & 0.5716{\scriptsize{$\pm$0.001}} & 0.4216{\scriptsize{$\pm$0.004}} & 0.7861{\scriptsize{$\pm$0.002}} & 0.6316{\scriptsize{$\pm$0.009}} & 0.5164{\scriptsize{$\pm$0.003}} \\
& SNR  & 4.2619{\scriptsize{$\pm$0.097}} & 2.3160{\scriptsize{$\pm$0.046}} & 1.0316{\scriptsize{$\pm$0.127}} & 4.7190{\scriptsize{$\pm$0.009}} & 2.6194{\scriptsize{$\pm$0.004}} & 1.6492{\scriptsize{$\pm$0.115}} \\
\midrule

\multirow{3}{*}{DDPMEEG}
& NMSE & 0.4916{\scriptsize{$\pm$0.004}} & 0.7319{\scriptsize{$\pm$0.002}} & 0.8634{\scriptsize{$\pm$0.007}} & 0.5136{\scriptsize{$\pm$0.004}} & 0.6513{\scriptsize{$\pm$0.001}} & 0.7916{\scriptsize{$\pm$0.005}} \\
& PCC  & 0.6941{\scriptsize{$\pm$0.002}} & 0.5134{\scriptsize{$\pm$0.001}} & 0.3419{\scriptsize{$\pm$0.004}} & 0.7346{\scriptsize{$\pm$0.001}} & 0.5316{\scriptsize{$\pm$0.009}} & 0.4305{\scriptsize{$\pm$0.003}} \\
& SNR  & 4.1943{\scriptsize{$\pm$0.095}} & 1.5391{\scriptsize{$\pm$0.042}} & 0.9431{\scriptsize{$\pm$0.125}} & 4.4165{\scriptsize{$\pm$0.008}} & 2.1064{\scriptsize{$\pm$0.003}} & 1.6105{\scriptsize{$\pm$0.110}} \\
\midrule

\multirow{3}{*}{ESTformer}
& NMSE & \underline{0.3288}{\scriptsize{$\pm$0.004}} & \underline{0.3483}{\scriptsize{$\pm$0.002}} & \underline{0.4149}{\scriptsize{$\pm$0.007}} & \underline{0.3448}{\scriptsize{$\pm$0.004}} & \underline{0.3911}{\scriptsize{$\pm$0.0015}} & \underline{0.5125}{\scriptsize{$\pm$0.005}} \\
& PCC  & \underline{0.8368}{\scriptsize{$\pm$0.002}} & \underline{0.8012}{\scriptsize{$\pm$0.001}} & \underline{0.7670}{\scriptsize{$\pm$0.004}} & \underline{0.8106}{\scriptsize{$\pm$0.002}} & \underline{0.7822}{\scriptsize{$\pm$0.009}} & \underline{0.7048}{\scriptsize{$\pm$0.003}} \\
& SNR  & \underline{5.0560}{\scriptsize{$\pm$0.097}} & \underline{4.5838}{\scriptsize{$\pm$0.044}} & \underline{3.8871}{\scriptsize{$\pm$0.126}} & \underline{4.7535}{\scriptsize{$\pm$0.008}} & \underline{4.1933}{\scriptsize{$\pm$0.003}} & \underline{2.9821}{\scriptsize{$\pm$0.113}} \\
\midrule

\multirow{3}{*}{STAD}
& NMSE & 0.4319{\scriptsize{$\pm$0.004}} & 0.6913{\scriptsize{$\pm$0.002}} & 0.8671{\scriptsize{$\pm$0.007}} & 0.3819{\scriptsize{$\pm$0.004}} & 0.6713{\scriptsize{$\pm$0.002}} & 0.7193{\scriptsize{$\pm$0.005}} \\
& PCC  & 0.7136{\scriptsize{$\pm$0.002}} & 0.4946{\scriptsize{$\pm$0.001}} & 0.3441{\scriptsize{$\pm$0.004}} & 0.7316{\scriptsize{$\pm$0.002}} & 0.5219{\scriptsize{$\pm$0.009}} & 0.4319{\scriptsize{$\pm$0.003}} \\
& SNR  & 4.1364{\scriptsize{$\pm$0.099}} & 1.4349{\scriptsize{$\pm$0.043}} & 0.9134{\scriptsize{$\pm$0.125}} & 4.4930{\scriptsize{$\pm$0.008}} & 2.0492{\scriptsize{$\pm$0.003}} & 1.6193{\scriptsize{$\pm$0.114}} \\
\midrule

\multirow{3}{*}{SRGDiff}
& NMSE & \textbf{0.1632}{\scriptsize{$\pm$0.004}} & \textbf{0.2977}{\scriptsize{$\pm$0.002}} & \textbf{0.3494}{\scriptsize{$\pm$0.007}} & \textbf{0.1663}{\scriptsize{$\pm$0.004}} & \textbf{0.2115}{\scriptsize{$\pm$0.002}} & \textbf{0.2603}{\scriptsize{$\pm$0.005}} \\
& PCC  & \textbf{0.9102}{\scriptsize{$\pm$0.002}} & \textbf{0.8445}{\scriptsize{$\pm$0.001}} & \textbf{0.8167}{\scriptsize{$\pm$0.004}} & \textbf{0.9113}{\scriptsize{$\pm$0.002}} & \textbf{0.8846}{\scriptsize{$\pm$0.009}} & \textbf{0.8210}{\scriptsize{$\pm$0.003}} \\
& SNR  & \textbf{7.8413}{\scriptsize{$\pm$0.097}} & \textbf{5.2606}{\scriptsize{$\pm$0.043}} & \textbf{4.5912}{\scriptsize{$\pm$0.127}} & \textbf{7.8660}{\scriptsize{$\pm$0.008}} & \textbf{6.6402}{\scriptsize{$\pm$0.003}} & \textbf{6.0346}{\scriptsize{$\pm$0.120}} \\

\bottomrule
\end{tabular}
}
\caption{Performance comparison of different models on SEED and SEED-IV datasets across different channel settings.}
\label{tab:SeedSR}
\end{table*}



\begin{figure}[]
  \centering
  \includegraphics[width=\columnwidth]{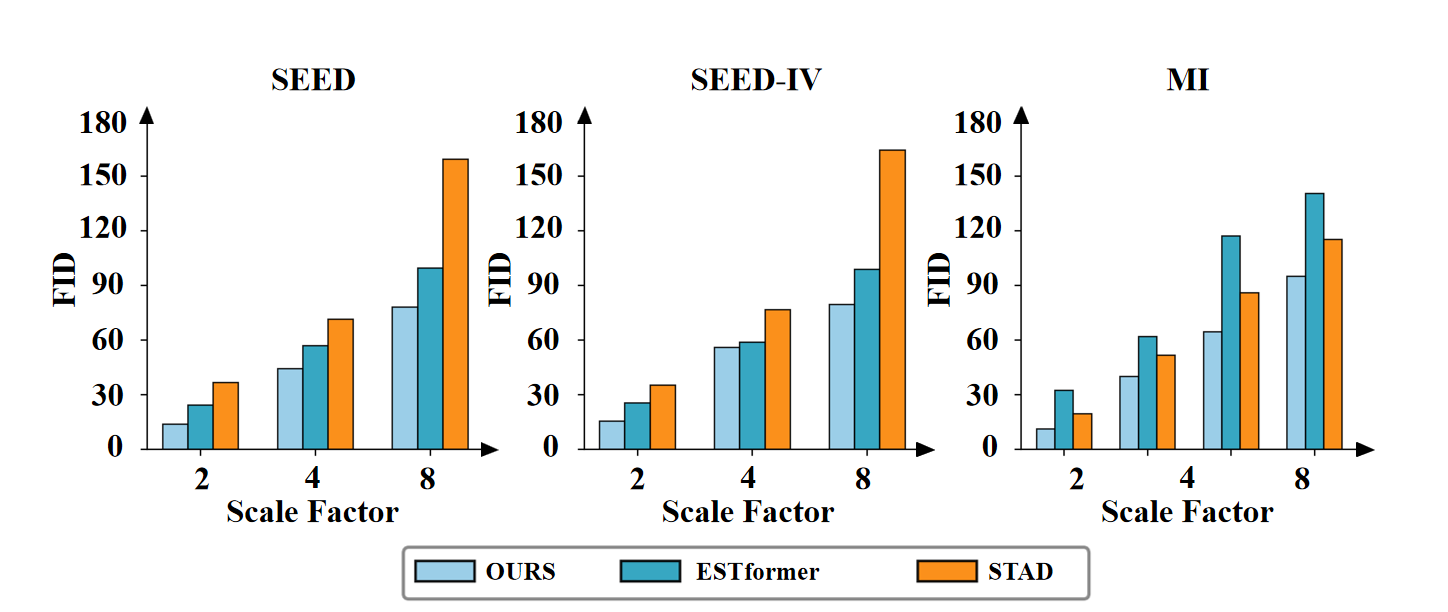}
  \caption{EEG-FID evaluation across three datasets compared with ESTformer and STAD.}
  \label{fig:fid_results}
\end{figure}

\subsection{Feature-level Reconstruction Evaluation}

We report EEG-FID results in Figure~\ref{fig:fid_results}, and our method consistently achieves the lowest FID scores across SEED, SEED-IV, and Localize-MI datasets under different scale factors. These results indicate that our approach generates EEG signals that are statistically closer to the real distribution in the temporal domain.

\begin{figure}[]
  \centering
    \includegraphics[width=\columnwidth]{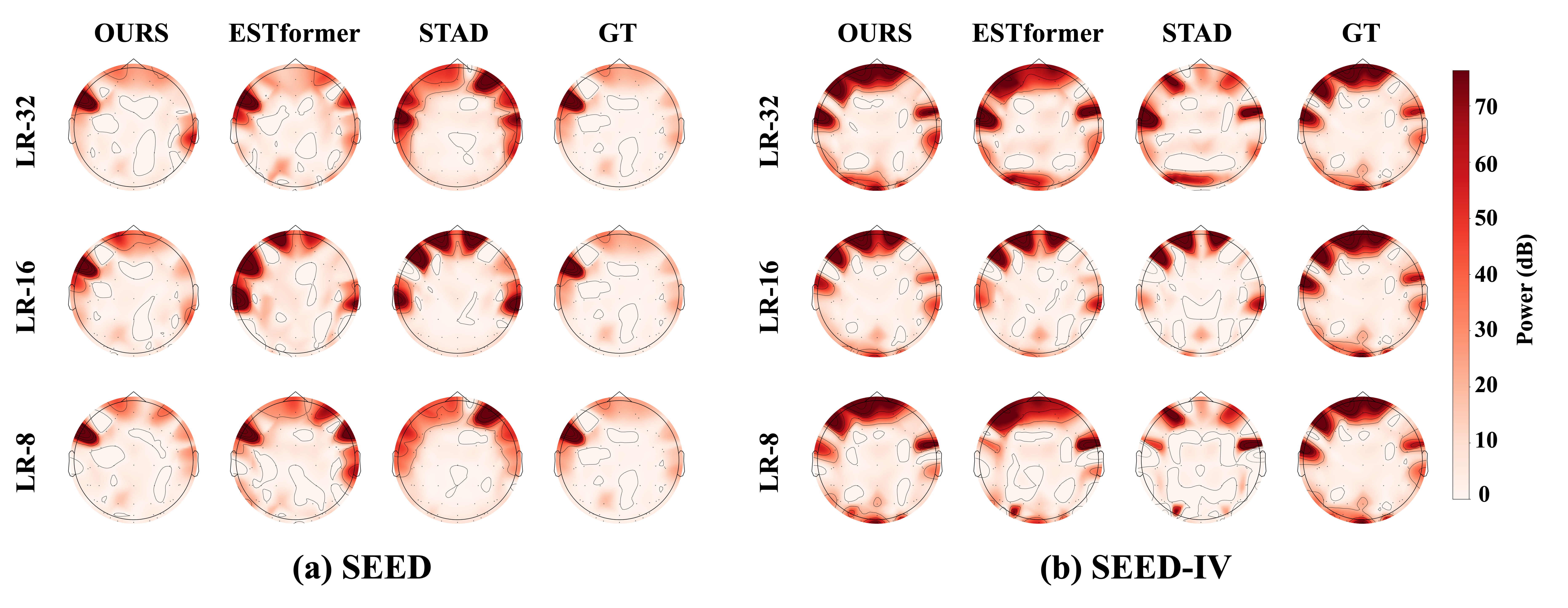}
  \caption{{Visualization of EEG topographic maps between ground-truth and reconstructed EEG signals by ESTformer, STAD and SRGDiff.}}
  \label{fig:topomap}
  
\end{figure}

We further analyze the spectral fidelity of generated signals by visualizing EEG topographic maps under different scale factors. An EEG topographic map projects the power spectral density (PSD) of each channel onto the scalp surface, providing an intuitive representation of the spatial distribution of oscillatory energy. As shown in Figure~\ref{fig:topomap} and Figure ~\ref{fig:topomap_MI}, although our reconstructed signals still exhibit minor deviations from the original data, they preserve a high degree of overlap in critical regions with strong PSD responses.

Beyond qualitative inspection, we also report a frequency-domain error metric that quantifies the mean absolute error between reconstructed and real HD topomaps. As shown in Table~\ref{tab:freq_topomap_mae}, SRGDiff consistently achieves the lowest frequency-domain MAE across datasets and SR scales, indicating better preservation of the spatial distribution of spectral power.

\begin{table}[t]
\centering
\small

{%
\begin{tabular}{lcccccccccc}
\toprule
\multirow{2}{*}{Model} &
\multicolumn{3}{c}{SEED} &
\multicolumn{3}{c}{SEED-IV} &
\multicolumn{4}{c}{Localize-MI} \\
\cmidrule(lr){2-4} \cmidrule(lr){5-7} \cmidrule(lr){8-11}
 & 2$\times$ & 4$\times$ & 8$\times$
 & 2$\times$ & 4$\times$ & 8$\times$
 & 2$\times$ & 4$\times$ & 8$\times$ & 16$\times$ \\
\midrule
ESTformer
 & 6.96  & 9.31  & 9.73
 & 7.11  & 8.30  & 8.86
 & 7.03  & 16.67 & 32.32 & 35.73 \\
STAD
 & 9.19  & 11.04 & 14.40
 & 9.50  & 10.95 & 13.12
 & 8.76  & 13.11 & 22.53 & 25.37 \\
\textbf{SRGDiff}
 & \textbf{3.89}  & \textbf{5.12}  & \textbf{4.95}
 & \textbf{3.99}  & \textbf{4.08}  & \textbf{4.84}
 & \textbf{3.86}  & \textbf{7.30}  & \textbf{11.50} & \textbf{13.76} \\
\bottomrule
\end{tabular}
}
\caption{Frequency-domain MAE between reconstructed and real HD topomaps on SEED, SEED-IV, and Localize-MI under different SR factors.}
\label{tab:freq_topomap_mae}

\end{table}

\subsection{Downstream tasks}

\begin{table*}[h]
  \centering
  \resizebox{\linewidth}{!}{
  \begin{tabular}{lcccccccccc}
    \toprule
    \multirow{2}{*}{\textbf{Method}} & \multicolumn{3}{c}{\textbf{SEED}} & \multicolumn{3}{c}{\textbf{SEED-IV}} & \multicolumn{4}{c}{\textbf{Localize-MI}} \\
    \cmidrule(lr){2-4} \cmidrule(lr){5-7} \cmidrule(lr){8-11}
    & 2 & 4 & 8 & 2 & 4 & 8 & 2 & 4 & 8 & 16 \\
    \midrule
    GT & 0.7152 & 0.7152 & 0.7152 & 0.7027 & 0.7027 & 0.7027 & 0.8368 & 0.8368 & 0.8368 & 0.8368 \\
    LR & 0.4981 & 0.4702 & 0.4424 & 0.5685 & 0.5618 & 0.5011 & 0.7208 & 0.6534 & 0.5219 & 0.3862 \\
    \midrule
    
    SaSDim & 0.5097 & 0.4793 & 0.4429 & 0.5794 & 0.5692 & 0.4912 & 0.7193 & 0.6519 & 0.5237 & 0.3845 \\
    
    SADI & 0.5137 & 0.4834 & 0.4456 & 0.5718 & 0.5644 & 0.4987 & 0.7215 & 0.6634 & 0.5314 & 0.3957 \\
    
    RDPI & 0.5044 & 0.4802 & 0.4531 & 0.5591 & 0.5741 & 0.5071 & 0.7230 & 0.6624 & 0.5210 & 0.3892 \\
    DDPMEEG & 0.4738 & 0.4610 & 0.4238 & 0.5548 & 0.5487 & 0.4838 & 0.7015 & 0.6387 & 0.5187 & 0.3767 \\
    ESTformer & \underline{0.6887} & \underline{0.6509} & \underline{0.6057} & \underline{0.6782} & \underline{0.6500} & \underline{0.5084} & 0.7445 & 0.6033 & 0.4739 & 0.4391 \\
    STAD & 0.5437 & 0.5249 & 0.4610 & 0.6651 & 0.6410 & 0.4977 & \underline{0.7589} & \underline{0.6797} & \underline{0.6344} & \underline{0.5384} \\
    SRGDiff & \textbf{0.7019} & \textbf{0.6812} & \textbf{0.6273} & \textbf{0.6821} & \textbf{0.6558} & \textbf{0.5127} & \textbf{0.7641} & \textbf{0.7163} & \textbf{0.6806} & \textbf{0.5887} \\
    \bottomrule
  \end{tabular}
  }
  \caption{Classification accuracy comparison across different methods and datasets. GT represents the ground truth performance. Best results are shown in bold, second-best are underlined.}
  \label{tab:classification}
\end{table*}

Table~\ref{tab:classification} reports results on the three datasets under various super-resolution scales.
As the scale grows, accuracy for both raw and super-resolved inputs declines, yet SRGDiff’s reconstructions consistently maintain a clear advantage. In particular, SRGDiff greatly outperforms missing-value imputation methods and leads all other spatial imputation approaches. At $2\times$ scale factor, its classification accuracy approaches that obtained from the original full-channel recordings.
We further compared the runtime efficiency of different methods as shown in Table~\ref{tab:runtime}. Although our proposed SRGDiff is slightly slower than the transformer-based ESTformer, it can still complete EEG super-resolution within 0.1s, which meets the real-time requirement in practical applications. The detailed runtime statistics of all models are provided in the Appendix.

\subsection{Ablation Study}

To evaluate the contribution of each module in SRGDiff to EEG super-resolution reconstruction, we conducted ablation studies comparing SRGDiff with three variant models. \textbf{LDM+LD} keeps only the VAE–DDIM backbone and takes the LD EEG as its input condition; \textbf{LDM+SMM} preserves the Step-aware modulation module; \textbf{LDM+RDM} retains the Residual Direction Module. All models were tested under the same experimental settings.

\begin{figure}[]
  \centering
  \includegraphics[width=0.95\linewidth]{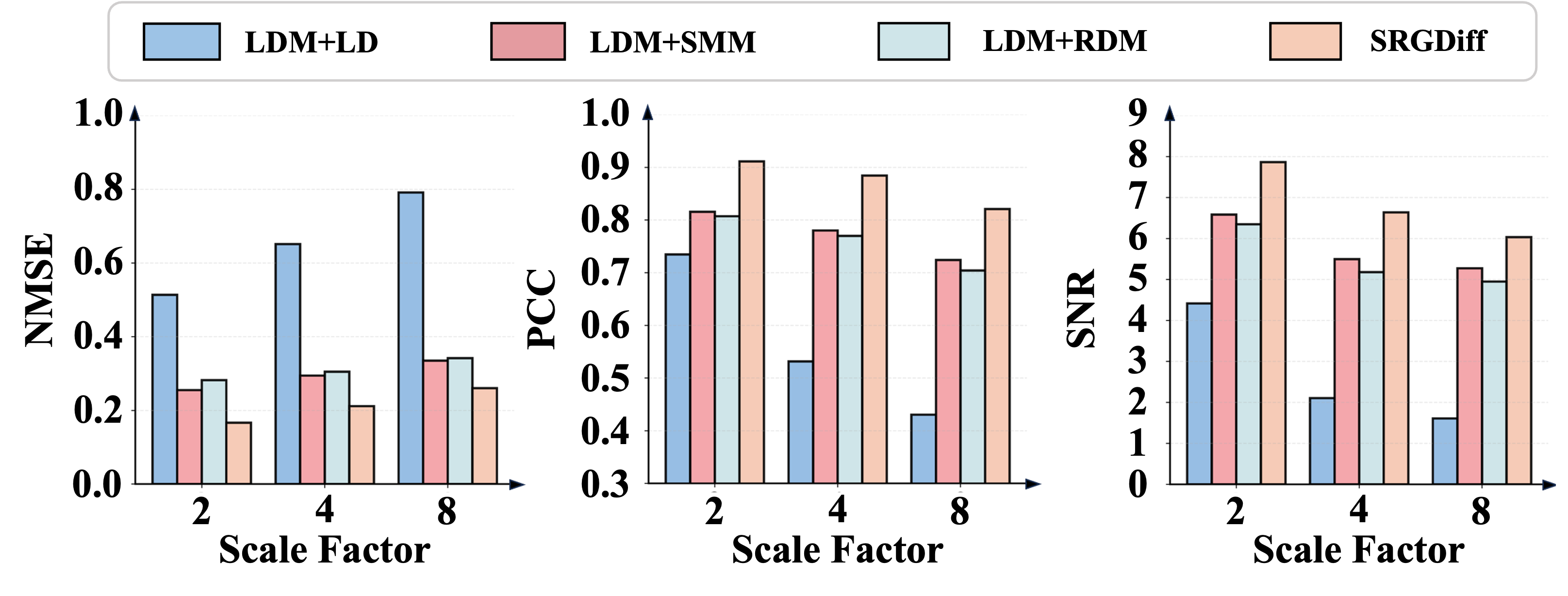}
  \caption{Ablation study performance comparison between SRGDiff and three variant models on the SEED dataset.}
  \label{fig:ablation}
\end{figure}

Figure~\ref{fig:ablation} reports NMSE, PCC and SNR across $2\times$, $4\times$ and $8\times$ upsampling in SEED dataset. At $8\times$, adding SMM to the baseline cuts NMSE from $0.86$ to $0.45$ with $47$\% reduction and boosts PCC from $0.34$ to $0.69$, demonstrating its effectiveness in temporally aligning the denoising trajectory. Incorporating RDM yields a comparable NMSE reduction with $44$\% and raises PCC to $0.67$, highlighting its role in injecting prior information for spatial consistency. When combined in SRGDiff, these modules further decrease NMSE to $0.34$ with $60$\% overall reduction and elevate PCC to $0.81$. More ablation results in SEED-IV and Localize-MI datasets are shown in Figure~\ref{fig:ablation_supp} in the Appendix.


\section{Conclusion}

We introduced SRGDiff, a step-aware residual-guided diffusion model that reframes EEG spatial super-resolution as guided HD generation with a step-aware residual direction and adaptive modulation. Across SEED, SEED-IV, and Localize-MI, SRGDiff consistently improves signal-level metrics, achieves the best EEG-FID across scales, and better preserves spectral and scalp topographies. Downstream evaluations further show higher accuracy on emotion recognition and patient-wise classification, indicating that the reconstructed signals are not only visually and statistically closer to HD EEG but also more useful for analysis. These results validate that explicit, step-wise conditioning on sparse inputs is both necessary and effective for high-fidelity, topology-preserving EEG super-resolution.

\section{Acknowledgements}

This work was supported by the National Key Research and Development Program of China (Grant No. 2022ZD0118001), the National Natural Science Foundation of China (Grant Nos. U22A2022, 62332017), and was conducted in part at the MOE Key Laboratory of Advanced Materials and Devices for Post-Moore Chips, the Beijing Key Laboratory of Big Data Innovation and Application for Skeletal Health Medical Care, and the Beijing Advanced Innovation Center for Materials Genome Engineering.

\section{Ethics Statement}
This work adheres to the ICLR Code of Ethics. In this study, no human subjects or animal experimentation was involved. All datasets used, including SEED, SEED-IV and Localize-MI, were sourced in compliance with relevant usage guidelines, ensuring no violation of privacy. Details could be found in Appendix. We have taken care to avoid any biases or discriminatory outcomes in our research process. No personally identifiable information was used, and no experiments were conducted that could raise privacy or security concerns. We are committed to maintaining transparency and integrity throughout the research process.

\section{Reproducibility Statement}
We have made every effort to ensure that the results presented in this paper are reproducible. All code and datasets have been made publicly available in an anonymous repository to facilitate replication and verification. The experimental setup, including training steps, model configurations, and hardware details, is described in detail in the paper. We have also provided a full description of SRGDiff, to assist others in reproducing our experiments.

We believe these measures will enable other researchers to reproduce our work and further advance the field.

\bibliography{iclr2026_conference}
\bibliographystyle{iclr2026_conference}

\appendix

\clearpage

\section{LLM Usage}
Large Language Models (LLMs) were used to aid in the writing and polishing of the manuscript. Specifically, we used an LLM to assist in refining the language, improving readability, and ensuring clarity in various sections of the paper. The model helped with tasks such as sentence rephrasing, grammar checking, and enhancing the overall flow of the text.

It is important to note that the LLM was not involved in the ideation, research methodology, or experimental design. All research concepts, ideas, and analyses were developed and conducted by the authors. The contributions of the LLM were solely focused on improving the linguistic quality of the paper, with no involvement in the scientific content or data analysis.

The authors take full responsibility for the content of the manuscript, including any text generated or polished by the LLM. We have ensured that the LLM-generated text adheres to ethical guidelines and does not contribute to plagiarism or scientific misconduct.

\begin{figure}[h]
  \centering
  \includegraphics[width=\columnwidth]{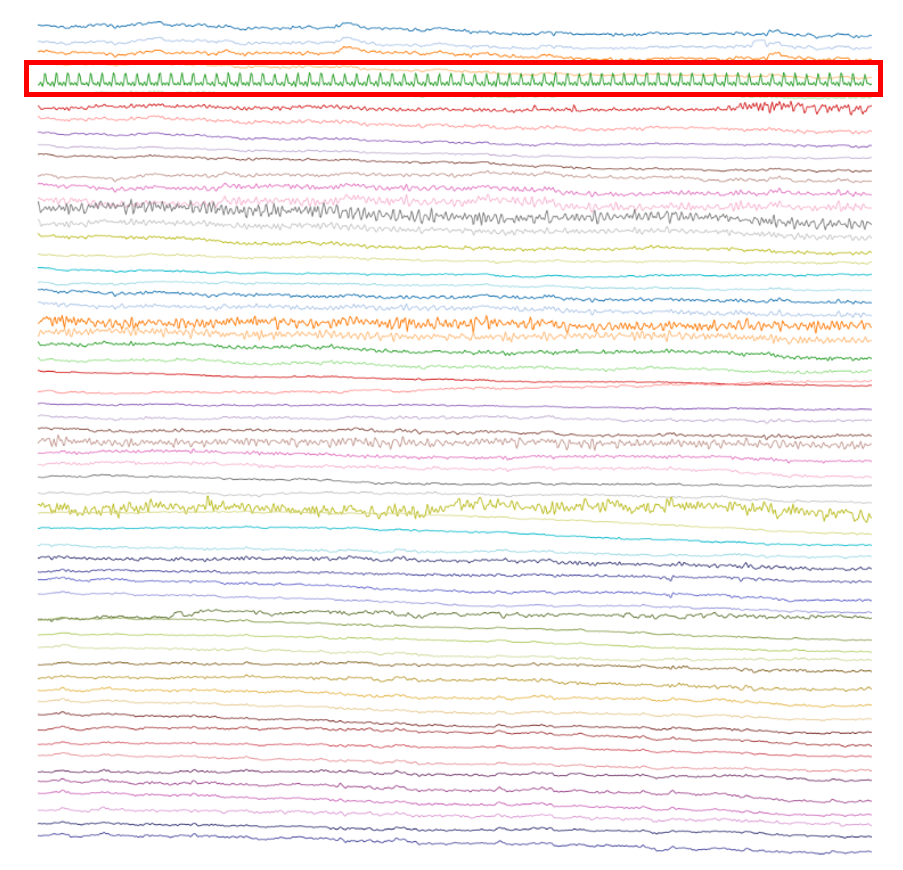}
  \caption{Example of a raw SEED EEG segment with sensor faults highlighted.}
  \label{fig:pre1}
\end{figure}

\begin{figure}[h]
  \centering
  \includegraphics[width=\columnwidth]{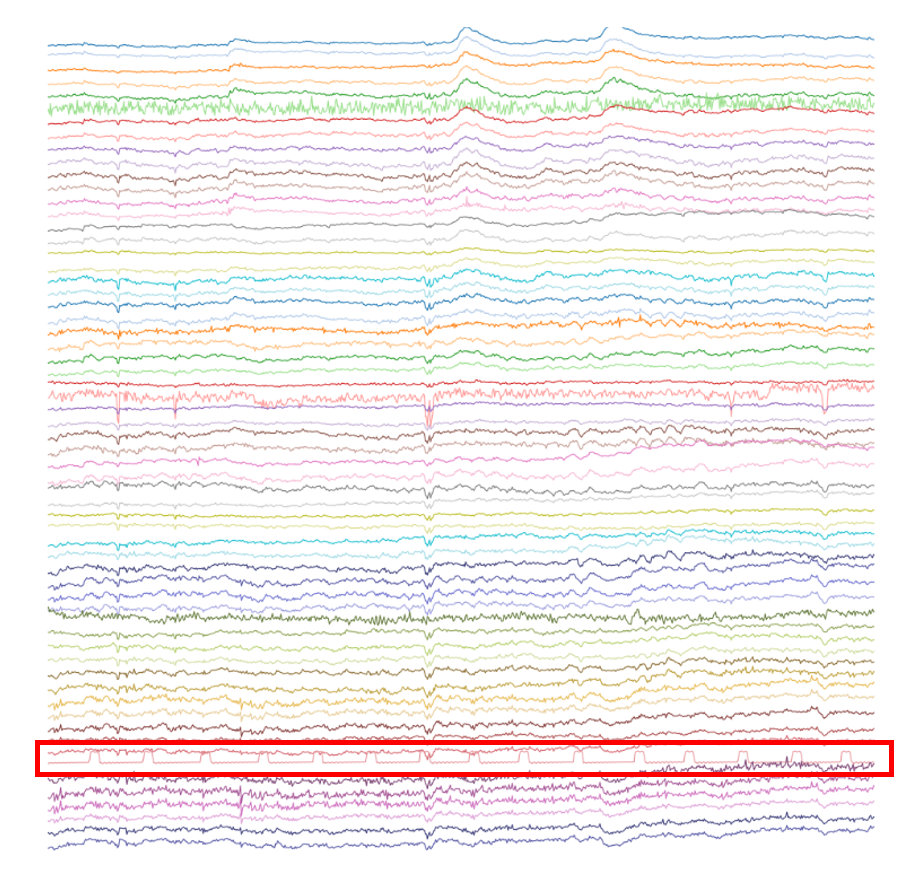}
  \caption{Example of a raw SEED EEG segment with sensor faults highlighted.}
  \label{fig:pre2}
\end{figure}

\section{Dataset Preprocess Details}

\subsection{Dataset Details}

We use three publicly available EEG datasets: SEED, SEED-IV, and Localize-MI.

\textbf{SEED} (available at \url{http://bcmi.sjtu.edu.cn/~seed/}) consists of recordings from $15$ subjects watching emotion-eliciting film clips ($\approx$ $4$ min each) designed to induce positive, neutral, and negative states. Data were acquired with a $62$-channel $10–20$ montage at $1000$ Hz, downsampled to $200$ Hz, and band-pass filtered to $0.5–75$ Hz. We manually inspected and removed sessions with sensor faults as depicted in Figure~\ref{fig:pre1} and Figure~\ref{fig:pre2}.

\textbf{SEED-IV} (available at \url{http://bcmi.sjtu.edu.cn/~seed/seed-iv.html}) is a publicly available EEG dataset designed for emotion recognition research. It includes four emotional categories: happiness, sadness, neutrality, and fear. Emotional states are elicited using two types of stimuli: music and images. EEG recordings were collected from 15 subjects using a 62-channel system with a sampling rate of 1000~Hz.During preprocessing, the EEG signals were downsampled to 200~Hz and filtered with a bandpass filter ranging from 0.5 to 75~Hz. To ensure data quality, visually corrupted or invalid trials were manually excluded. 

\textbf{Localize-MI} (available at \url{https://doi.org/10.12751/g-node.1cc1ae}) is a high-density intracranial EEG dataset from seven drug-resistant epilepsy patients during $61$ presurgical sessions. Stereo–EEG electrodes delivered single-pulse biphasic currents ($0.1–5$ mA), and $256$ channels were recorded at $8000$ Hz. Preprocessing included $0.1$ Hz high-pass filtering, notch filters at $50$/$100$/$150$/$200$ Hz, bad-channel/trial removal, and trial alignment using stimulation artifact peaks ($–300$ to $+50$ ms window).
In the Localize-MI dataset, we designed a binary classification task (epileptic vs. nonepileptic) to evaluate the effectiveness of synthetic super-resolution EEG (SR EEG) in detecting epileptic abnormalities. Specifically, EEG signals recorded before electrical stimulation are labeled as nonepileptic, while those recorded during stimulation are labeled as epileptic. The detailed experimental setup follows the description provided in the STAD \citep{wang2025generative} model section.

\subsection{More Experimental Details}

\begin{figure}[h]
  \centering
  \includegraphics[width=\columnwidth]{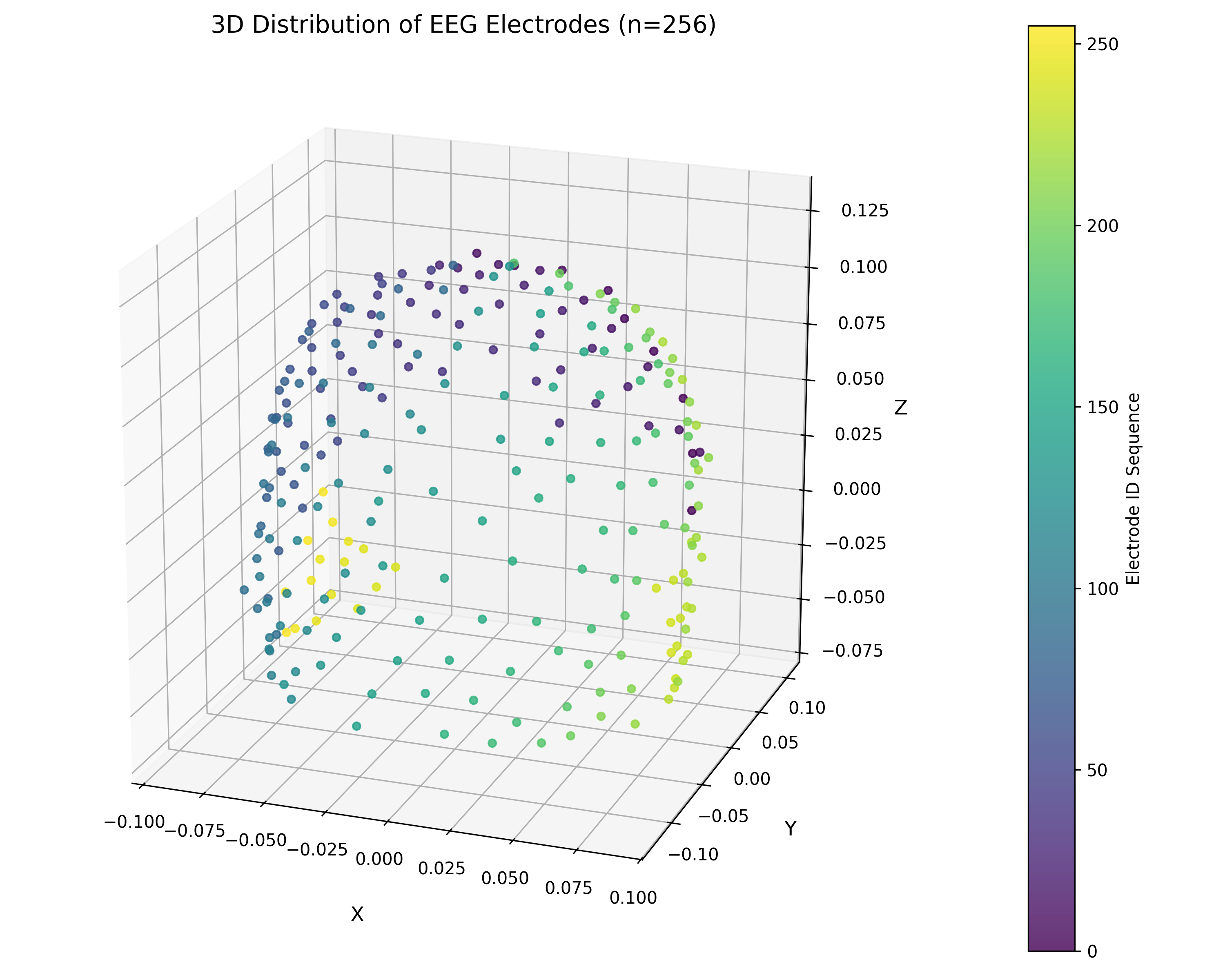}
  \caption{Electrode topology of the Localize-MI dataset (256 intracranial channels).}
  \label{fig:topology_mi}
\end{figure}

\begin{figure}[h]
  \centering
  \includegraphics[width=\columnwidth]{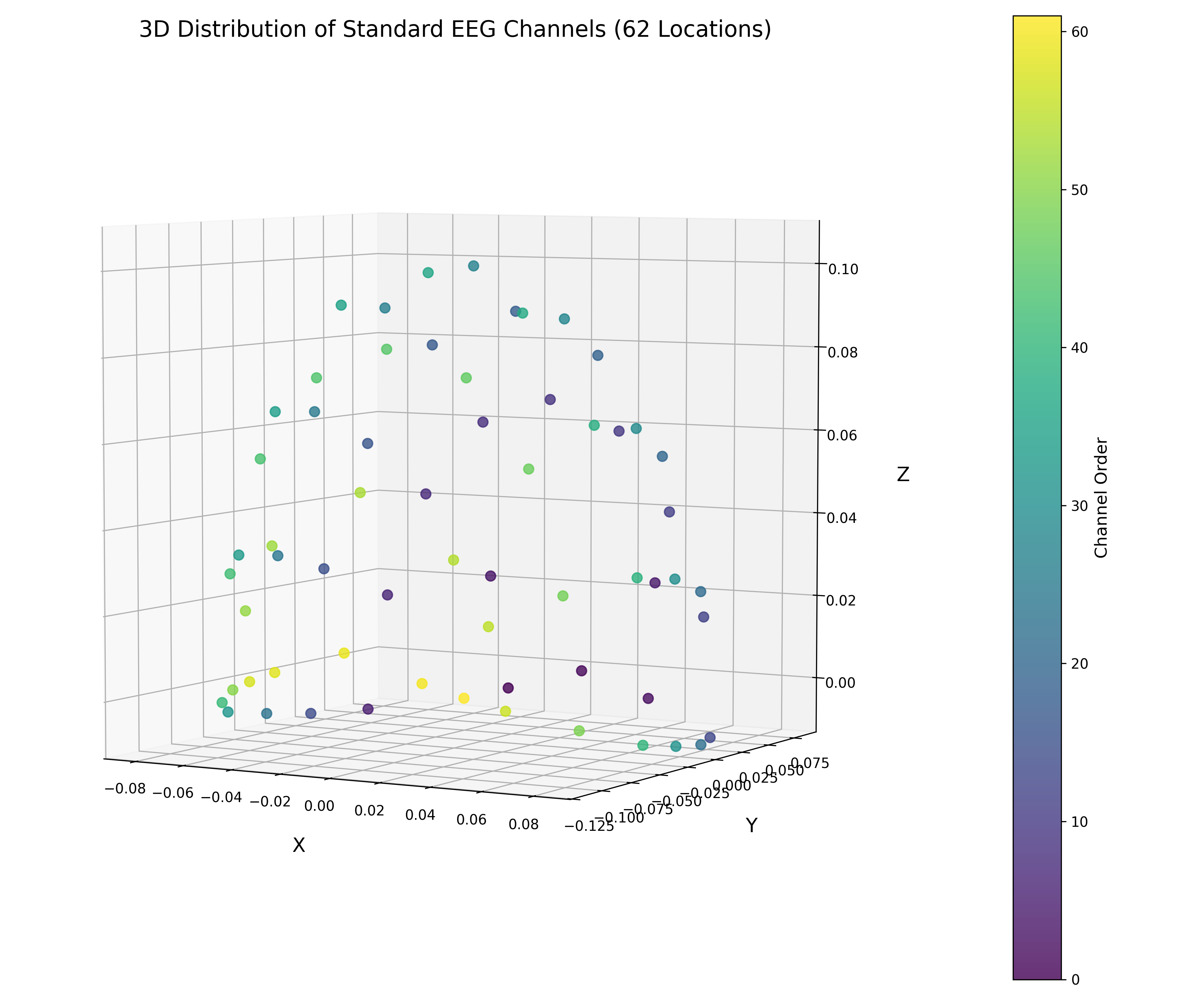}
  \caption{Electrode topology of the SEED-IV and SEED datasets (62 scalp channels).}
  \label{fig:topology_seed}
\end{figure}

We follow ESTformer and STAD slicing strategies. Preprocessed signals are windowed into fixed lengths: SEED and SEED-IV use non-overlapping $4$ s segments, while Localize-MI retains $–250$ ms to $+10$ ms around each stimulus ($260$ ms total). We randomly split $80$\% for training and $20$\% for testing, yielding $24265\times62\times800$ train / $6067\times62\times800$ test samples for SEED; $29199\times62\times800$ train / $7300\times62\times800$ test for SEED-IV; and $1914\times256\times2081$ train / $479\times256\times2081$ test for Localize-MI.  
For SEED and SEED-IV we evaluate $2\times$, $4\times$, and $8\times$ super-resolution; for Localize-MI we additionally include $16\times$.
As shown in Figure~\ref{fig:topology_mi}, Localize-MI employs a 256-channel intracranial grid, while Figure~\ref{fig:topology_seed} shows the 62-channel scalp montage used in SEED-IV and SEED.

\section{More SRGDiff Model Details}

\subsection{Variational Autoencoder}

In this paper, Variational Autoencoder (VAE) follows the AutoencoderKL design {\citep{aristimunha2023synthetic}}, comprising a convolutional encoder, a latent distribution (mean and variance) with KL regularization toward \(\mathcal{N}(0,I)\), and a decoder with deconvolutions and upsampling. We augment both encoder and decoder with attention layers (multi-head and non-local attention), residual connections, and GroupNorm to capture global EEG features while ensuring stable training and efficient latent representations.

\subsection{DDIM Scheduler}

The DDIMScheduler manages noise scheduling and sampling in the forward and reverse diffusion processes. It supports multiple noise prediction types and variance strategies. At each step, it computes the noise coefficient, predicts the denoised sample, clips values for numerical stability, and injects random perturbations to control output diversity.

\section{Parameter Study}

\subsection{VAE Latent Shape Selection}

We found that the latent shape balances reconstruction precision and generalization. Higher dimensions capture more detail but risk overfitting, while lower dimensions blur outputs. We experimented across the three datasets and selected a latent of \(32\times400\) for SEED/SEED-IV and \(64\times500\) for Localize-MI, which yielded optimal NMSE, PCC, and SNR. Table~\ref{tab:vae_shape_seed_seediv} and Table~\ref{tab:localizemi_shape} shows the performance of SRGDiff in different latent shapes.

\begin{table}[t]
  \centering
  \begin{minipage}{0.47\textwidth}
    \centering
    \begin{tabular}{lccc}
      \toprule
      \textbf{Shape} & \textbf{NMSE} & \textbf{PCC} & \textbf{SNR (dB)} \\
      \midrule
      $64\times400$        & 0.15 & 0.93 & 7.75 \\
      \textbf{32$\times$400} & \textbf{0.12} & \textbf{0.95} & \textbf{8.72} \\
      $16\times200$        & 0.20 & 0.89 & 6.81 \\
      $8\times400$         & 0.16 & 0.92 & 7.26 \\
      \bottomrule
    \end{tabular}
    \caption*{SEED}
  \end{minipage}
  \hfill
  \begin{minipage}{0.47\textwidth}
    \centering
    \begin{tabular}{lccc}
      \toprule
      \textbf{Shape} & \textbf{NMSE} & \textbf{PCC} & \textbf{SNR (dB)} \\
      \midrule
      $64\times400$        & 0.16 & 0.92 & 7.58 \\
      \textbf{32$\times$400} & \textbf{0.13} & \textbf{0.94} & \textbf{8.58} \\
      $16\times200$        & 0.21 & 0.87 & 6.86 \\
      $8\times400$         & 0.19 & 0.91 & 7.23 \\
      \bottomrule
    \end{tabular}
    \caption*{SEED-IV}
  \end{minipage}

  \caption{VAE latent shape selection results on SEED and SEED-IV.}
  \label{tab:vae_shape_seed_seediv}
\end{table}

\begin{table}[h]
  \centering
  \begin{tabular}{lccc}
    \toprule
    \textbf{Shape} & \textbf{NMSE} & \textbf{PCC} & \textbf{SNR (dB)} \\
    \midrule
    $128\times1000$         & 0.13 & 0.94 & 8.60 \\
    \textbf{64$\times$500}   & \textbf{0.09} & \textbf{0.96} & \textbf{9.01} \\
    $32\times500$           & 0.15 & 0.92 & 7.61 \\
    $32\times1000$          & 0.13 & 0.94 & 8.62 \\
    \bottomrule
  \end{tabular}
    \caption{Localize-MI: VAE latent shape selection results}
  \label{tab:localizemi_shape}
\end{table}

\subsection{Diffusion Hyperparameters}

\subsubsection{Diffusion Schedules} 
We compare linear and cosine noise schedules. Linear adds noise at a constant rate but may cause instability at endpoints; cosine offers smoother transitions and better performance for long diffusion chains. We fixed $1000$ timesteps with cosine scheduling and evaluated NMSE, PCC, and SNR on the latent reconstructions to choose this setting as shown in Table~\ref{tab:noise_schedule}.

\begin{table}[h]
  \centering
  \begin{tabular}{llccc}
    \toprule
    \textbf{Dataset} & \textbf{Schedule} & \textbf{NMSE} & \textbf{PCC} & \textbf{SNR (dB)} \\
    \midrule
    \multirow{2}{*}{SEED}    & Linear & 0.42 & 0.71 & 4.18 \\
                             & \textbf{Cosine} & \textbf{0.20} & \textbf{0.86} & \textbf{7.15} \\
    \midrule
    \multirow{2}{*}{SEED-IV} & Linear & 0.51 & 0.66 & 4.02 \\
                             & \textbf{Cosine} & \textbf{0.19} & \textbf{0.88} & \textbf{7.24} \\
    \midrule
    \multirow{2}{*}{Localize-MI} & Linear & 0.14 & 0.93 & 8.39 \\
                                 & \textbf{Cosine} & \textbf{0.11} & \textbf{0.95} & \textbf{8.88} \\
    \bottomrule
  \end{tabular}
    \caption{Comparison of noise schedules on three datasets (NMSE, PCC, SNR).}
  \label{tab:noise_schedule}
\end{table}

\subsubsection{Training Timestep Lengths} 
We also tested different training timestep lengths ($200$, $1000$, $2000$). Larger values introduce stronger noise but make denoising harder; smaller values lack coverage of high-noise regimes. Using cosine scheduling, the results in Table~\ref{tab:timesteps} exhibit that $1000$ timesteps to be optimal across datasets.

\begin{table}[h]
  \centering
  \begin{tabular}{llccc}
    \toprule
    \textbf{Dataset}   & \textbf{Steps}      & \textbf{NMSE} & \textbf{PCC} & \textbf{SNR (dB)} \\
    \midrule
    \multirow{3}{*}{SEED}      & 200                & 0.32          & 0.76         & 5.95            \\
                               & \textbf{1000}        & \textbf{0.20}          & \textbf{0.86}         & \textbf{7.15}            \\
                               & 2000               & 0.29          & 0.78         & 6.19            \\
    \midrule
    \multirow{3}{*}{SEED-IV}   & 200                & 0.36          & 0.75         & 5.93            \\
                               & \textbf{1000}        & \textbf{0.19}          & \textbf{0.88}         & \textbf{7.24}            \\
                               & 2000               & 0.31          & 0.76         & 6.11            \\
    \midrule
    \multirow{3}{*}{Localize-MI}& 200                & 0.15          & 0.92         & 8.37            \\
                               & 1000               & 0.11          & 0.95         & 8.88            \\
                               & \textbf{2000}        & \textbf{0.11}          & \textbf{0.95}         & \textbf{8.94}            \\
    \bottomrule
  \end{tabular}
    \caption{Impact of training timesteps on reconstruction quality}
  \label{tab:timesteps}
\end{table}

\subsubsection{Cosine Schedule Offset Factor}

In the cosine noise schedule for DDIM, the offset factor \(s\) adjusts the smoothness and starting point of the noise variance curve to prevent instability from overly small initial noise levels. Concretely, \(s\) introduces a phase shift in the cosine function, producing a more gradual noise increase at early timesteps—thereby avoiding abrupt noise jumps—while still covering the full variance range at later steps. Smaller values of \(s\) yield gentler initial noise ramp-up, whereas larger \(s\) accelerate early noise growth.  Table~\ref{tab:cos_offset} depicts the effect of cosine schedule offset factors on reconstruction quality.

\begin{table}[]
  \centering
  \begin{tabular}{lcccc}
    \toprule
    \textbf{Dataset}     & \(\boldsymbol{s}\) & \textbf{NMSE} & \textbf{PCC} & \textbf{SNR (dB)} \\
    \midrule
    \multirow{3}{*}{SEED}      & \textbf{0.005}  & \textbf{0.20} & \textbf{0.86} & \textbf{7.15} \\
                               & 0.010         & 0.24 & 0.84 & 7.02 \\
                               & 0.025         & 0.26 & 0.83 & 6.93 \\
    \midrule
    \multirow{3}{*}{SEED-IV}   & 0.005         & 0.20 & 0.86 & 7.17 \\
                               & \textbf{0.010}  & \textbf{0.19} & \textbf{0.88} & \textbf{7.24} \\
                               & 0.025         & 0.20 & 0.85 & 7.11 \\
    \midrule
    \multirow{3}{*}{Localize-MI}& \textbf{0.005}  & \textbf{0.11} & \textbf{0.95} & \textbf{8.94} \\
                               & 0.010         & 0.15 & 0.93 & 8.41 \\
                               & 0.025         & 0.18 & 0.91 & 8.23 \\
    \bottomrule
  \end{tabular}
    \caption{Effect of cosine schedule offset factor $s$ on reconstruction quality}
  \label{tab:cos_offset}
\end{table}

{\subsection{Effect of $\lambda_{\text{res}}$ and $\lambda_{\text{SMM}}$.}}

{
To assess the sensitivity of SRGDiff to the weighting coefficients in the loss, we conduct a parameter study on the most challenging SR settings (highest SR factor) for each dataset. We vary the residual-guidance weight $\lambda_{\text{res}}$ and the step-aware modulation weight $\lambda_{\text{SMM}}$ around the default values used in the main paper, while keeping all other hyperparameters fixed.
}

{
Concretely, we sweep
$\lambda_{\text{res}} \in \{0.1, 0.5, 1.0, 2.0, 5.0\}$ (relative to the default), and
$\lambda_{\text{SMM}} \in \{0.001, 0.005, 0.01, 0.02, 0.1\}$.
Tables~\ref{tab:param_lambda_both} report the NMSE on SEED, SEED-IV, and Localize-MI under these settings.}

\begin{table}[ht]
\centering

{%
\begin{minipage}{0.47\textwidth}
  \centering
  \begin{tabular}{lccc}
    \toprule
    $\lambda_{\text{res}}$ & SEED & SEED-IV & Localize-MI \\
    \midrule
    0.1 & 0.3928 & 0.3286 & 0.3941 \\
    0.5 & 0.3532 & 0.2822 & 0.3578 \\
    \textbf{1.0} & \textbf{0.3494} & \textbf{0.2603} & \textbf{0.3457} \\
    2.0 & 0.3508 & 0.2810 & 0.3565 \\
    5.0 & 0.4012 & 0.3369 & 0.3827 \\
    \bottomrule
  \end{tabular}
  \caption*{(a) Effect of $\lambda_{\text{res}}$}
\end{minipage}
\hfill
\begin{minipage}{0.47\textwidth}
  \centering
  \begin{tabular}{lccc}
    \toprule
    $\lambda_{\text{SMM}}$ & SEED & SEED-IV & Localize-MI \\
    \midrule
    0.001 & 0.3975 & 0.3133 & 0.3904 \\
    0.005 & 0.3539 & 0.2696 & 0.3519 \\
    \textbf{0.01} & \textbf{0.3494} & \textbf{0.2603} & \textbf{0.3457} \\
    0.02 & 0.3513 & 0.2712 & 0.3489 \\
    0.1  & 0.3990 & 0.3240 & 0.3915 \\
    \bottomrule
  \end{tabular}
  \caption*{(b) Effect of $\lambda_{\text{SMM}}$}
\end{minipage}
}%
\caption{Effect of $\lambda_{\text{res}}$ and $\lambda_{\text{SMM}}$ on NMSE (hardest SR setting per dataset).}
\label{tab:param_lambda_both}

\end{table}

{
Overall, the performance is reasonably stable within a broad range around the default values, indicating that SRGDiff is not overly sensitive to these hyperparameters. When $\lambda_{\text{res}}$ becomes too large, the residual guidance term dominates and suppresses learning in the diffusion backbone; when it is too small, the residual guidance has almost no effect. Similarly, if $\lambda_{\text{SMM}}$ is too small, the guidance feature modulation is overly strong, whereas for very large $\lambda_{\text{SMM}}$ the diffusion model effectively ignores the guidance features. The default configuration achieves the best overall trade-off across datasets.
}

\section{Downstream Task}

\subsection{Classification Feature Extraction}

\subsubsection{Differential Entropy Feature}

Raw EEG at $1000$ Hz is downsampled to $200$ Hz, band-pass filtered ($1–50$ Hz) with a $6$th-order Butterworth filter, and segmented into non-overlapping $1$ s windows ($200$ samples). Each window is transformed by STFT (Hanning window, $200$-point length, 256-point FFT). We compute band-specific power \(E\) for \(\delta\,(1\text{--}3\,\mathrm{Hz})\), \(\theta\,(4\text{--}7\,\mathrm{Hz})\), \(\alpha\,(8\text{--}13\,\mathrm{Hz})\), \(\beta\,(14\text{--}30\,\mathrm{Hz})\), and \(\gamma\,(31\text{--}50\,\mathrm{Hz})\)
, normalize by the number of bins \(N\), and define differential entropy (DE) feature as \(\log(E/N)\) with a small constant added for numerical stability.

\subsubsection{Power Spectral Density Feature}

Power spectral density (PSD) feature features use the same STFT pipeline but report the mean squared magnitude (average power) in each band.

\subsection{Random Forest Classifier}

For emotion classification on SEED and SEED-IV and epileptic detection on Localize-MI, we employ a random forest with 100 trees. 
This set of hyperparameters balances nonlinearity modeling with computational efficiency, yielding robust performance on high-dimensional EEG features.

\section{Supplemental Ablation Results}

\subsection{Frequency-domain Topomap Error}

{To complement the qualitative topographic visualizations in Figure~\ref{fig:topomap} and rule out potential visual selection bias, we report a quantitative frequency-domain MAE between reconstructed and real HD topomaps. Concretely, we first transform both reconstructed and reference HD EEG into the frequency domain, aggregate power within standard EEG bands (e.g., $\theta$, $\alpha$, $\beta$), and interpolate the band power of each channel onto a 2D scalp grid using electrode coordinates. We then compute the pixel-wise normalized mean squared error between the reconstructed and real topomaps, averaged over all frequency bands and test samples. A lower value indicates that the model better preserves both the spectral content and its spatial distribution over the scalp.}

{Table~\ref{tab:freq_topomap_nmse} reports frequency-domain MAE on SEED, SEED-IV, and Localize-MI under different SR factors. SRGDiff consistently achieves the lowest error across all datasets and scales, with a larger margin over ESTformer and STAD than in time-domain NMSE/PCC/SNR. This confirms that our residual-guided generative formulation not only improves pointwise reconstruction quality, but also more faithfully recovers the HD spectral–spatial structure.}
\begin{table}[t]
\centering
\small
{%
\begin{tabular}{lcccccccccc}
\toprule
\multirow{2}{*}{Model} &
\multicolumn{3}{c}{SEED} &
\multicolumn{3}{c}{SEED-IV} &
\multicolumn{4}{c}{Localize-MI} \\
\cmidrule(lr){2-4} \cmidrule(lr){5-7} \cmidrule(lr){8-11}
 & 2$\times$ & 4$\times$ & 8$\times$
 & 2$\times$ & 4$\times$ & 8$\times$
 & 2$\times$ & 4$\times$ & 8$\times$ & 16$\times$ \\
\midrule
ESTformer
 & 6.96  & 9.31  & 9.73
 & 7.11  & 8.30  & 8.86
 & 7.03  & 16.67 & 32.32 & 35.73 \\
STAD
 & 9.19  & 11.04 & 14.40
 & 9.50  & 10.95 & 13.12
 & 8.76  & 13.11 & 22.53 & 25.37 \\
\textbf{SRGDiff}
 & \textbf{3.89}  & \textbf{5.12}  & \textbf{4.95}
 & \textbf{3.99}  & \textbf{4.08}  & \textbf{4.84}
 & \textbf{3.86}  & \textbf{7.30}  & \textbf{11.50} & \textbf{13.76} \\
\bottomrule
\end{tabular}
\caption{{Frequency-domain NMSE between reconstructed and real HD topomaps on SEED, SEED-IV, and Localize-MI under different SR factors.}}
\label{tab:freq_topomap_nmse}
}%

\end{table}

\subsection{Additional Ablation Studies on SEED-IV and Localize-MI datasets}

\begin{figure}[h]
  \centering
    \includegraphics[width=0.9\columnwidth]{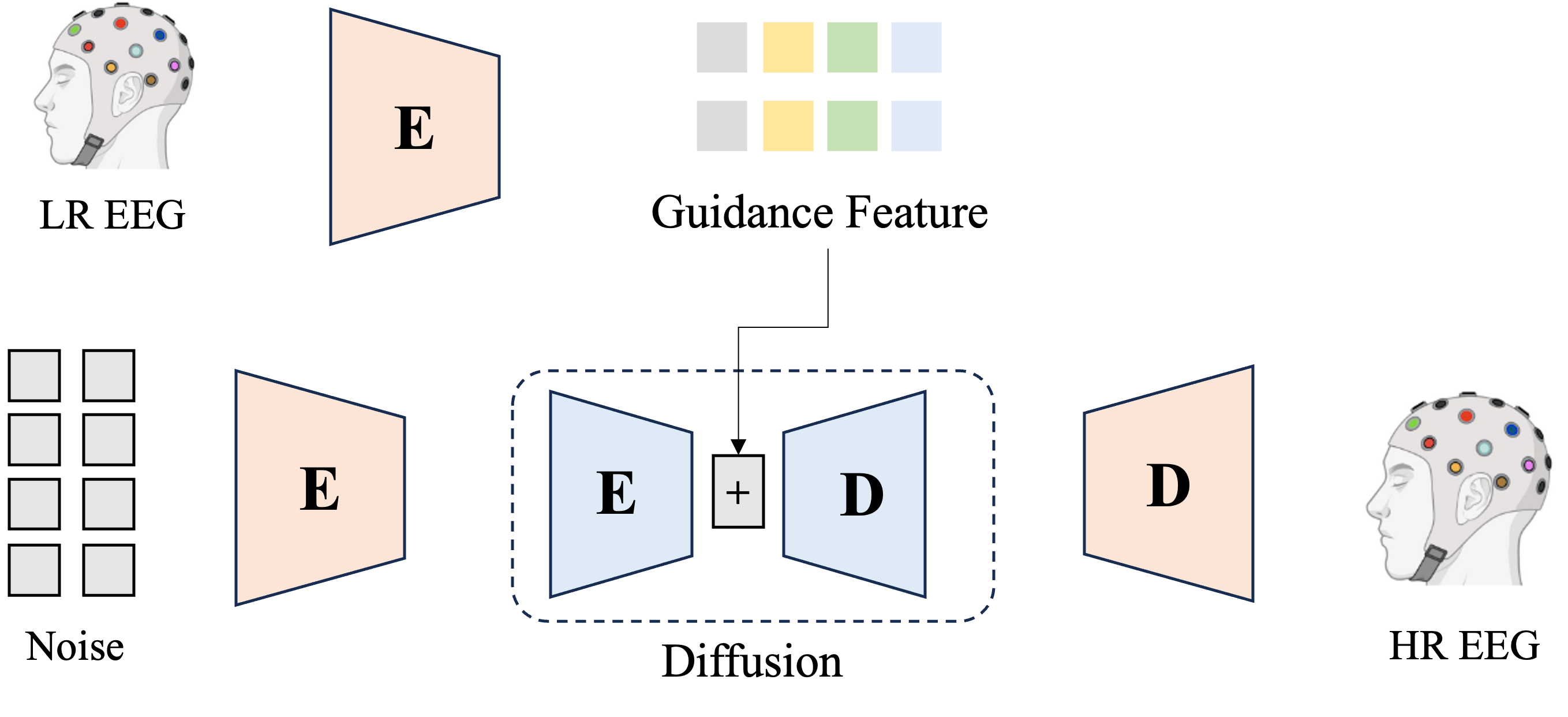}
  \caption{{Illustration of the LDM+LD baseline.}}
  \label{fig:ablation_baseline}
  
\end{figure}

{Figure~\ref{fig:ablation_baseline} illustrates the \textbf{LDM+LD} baseline used in our ablations. 
In the first stage (top row), we pretrain a VAE on full high-density (HD) EEG: preprocessed HD signals are passed through the encoder $E$ and decoder $D$ to learn a latent space tailored to HD scalp topography.
In the second stage (bottom row), low-density EEG is first mapped into this latent space using the pretrained encoder $E$, yielding a static guidance feature. This guidance feature is then added to the diffusion latent during the denoising process, and the final denoised latent is decoded by $D$ back to HD EEG, without employing RDM or SMM.}

\begin{figure}[h]
  \centering
  \includegraphics[width=0.9\columnwidth]{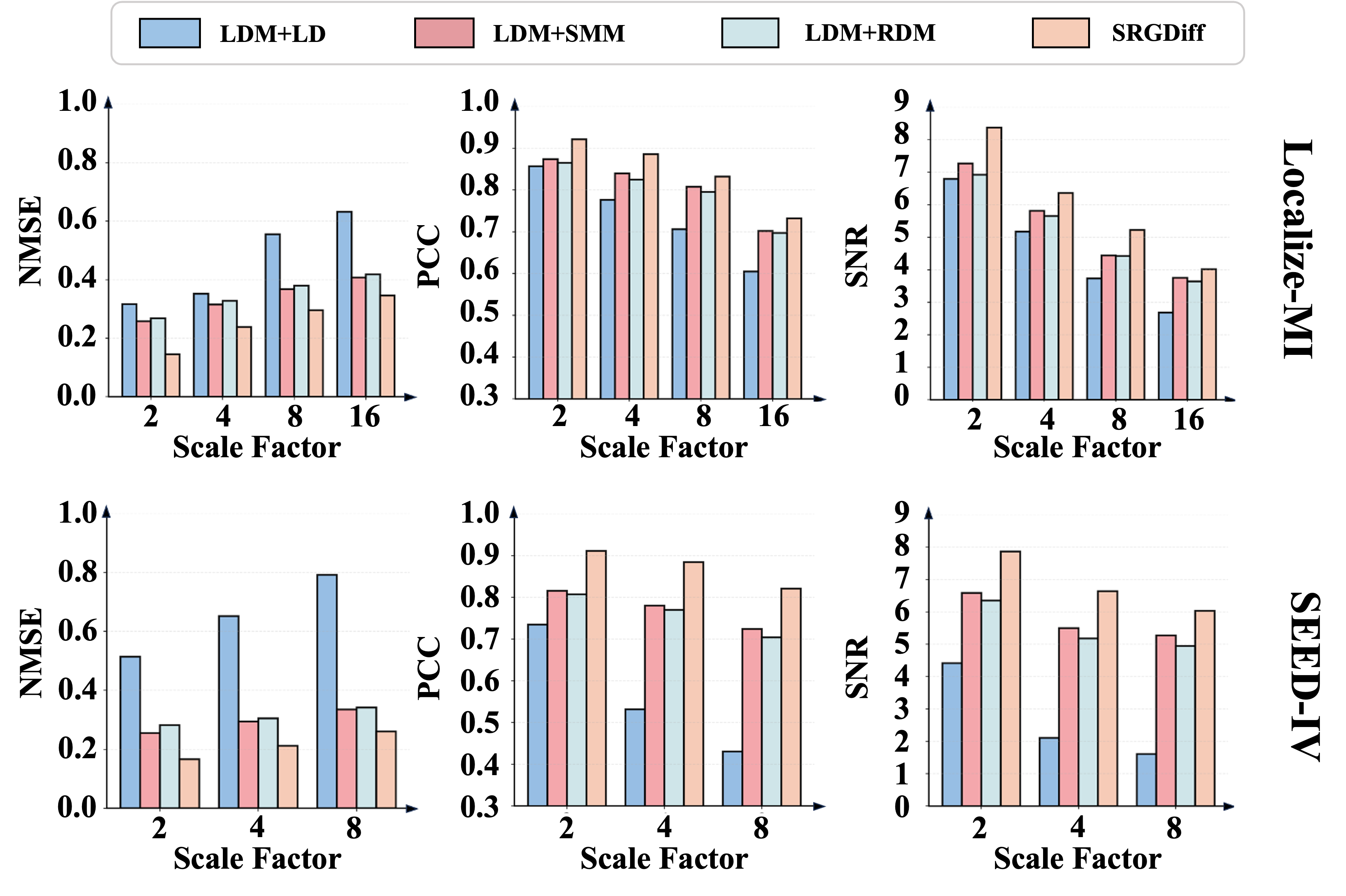}
  \caption{Ablation results on SEED-IV and Localize-MI: comparison of NMSE, PCC, and SNR for LDM+LD, LDM+SMM, LDM+RDM, and SRGDiff at 2×, 4×, and 8× upsampling and 2×, 4×, 8×, and 16× upsampling, respectively.}
  \label{fig:ablation_supp}
\end{figure}

Figure~\ref{fig:ablation_supp} presents additional ablation studies on SEED-IV and Localize-MI datasets, reporting NMSE, PCC, and SNR for the baseline LDM+LD, LDM+SMM, LDM+RDM, and the full SRGDiff across various upsampling scales. These plots further illustrate the individual and combined contributions of our two modules to reconstruction quality.

\section{Efficiency Analysis}

\subsection{Computational Cost under Comparable Parameter Budgets}

To complement the main-paper results, we further compare the computational cost of SRGDiff with simpler transformer-based SR models under comparable parameter budgets. Table~\ref{tab:comp_cost} reports, for ESTformer, STAD, and SRGDiff, the total number of trainable parameters and the average computation cost per 4\,s EEG window (measured as FLOPs under the same input resolution). Although SRGDiff requires more FLOPs than the single-pass transformer ESTformer due to the iterative denoising process, its parameter count remains in the same order of magnitude as transformer-based baselines, and the additional cost is the price we pay for exploiting a strong latent diffusion prior.

\begin{table}[hb]
\centering
{%
\small
\begin{tabular}{lcc}
\toprule
\textbf{Method} & \textbf{\#Params (M)} & \textbf{GFLOPs / window} \\
\midrule
ESTformer & 12.111 & 4.302 \\ 
STAD      & 13.949 & 1.650 \\ 
SRGDiff   & 2.342 & 1.38 \\ 
\bottomrule
\end{tabular}
\caption{Computational cost of different SR models under comparable parameter budgets.
We report the total number of trainable parameters and GFLOPs per 4\,s EEG window.
SRGDiff uses a diffusion-based denoiser, so its FLOPs are higher than those of ESTformer,
but the parameter budget remains comparable.}
\label{tab:comp_cost}
}%

\end{table}

\subsection{Runtime Comparison}

In addition to static computational cost, we also measure wall-clock runtime for all methods
on the three EEG datasets under a unified implementation and hardware setup (same GPU,
batch size, and input length). Table~\ref{tab:runtime} reports the average per-sample latency
for a 4\,s window.
Despite using an iterative denoising process, SRGDiff achieves inference times between
63\,ms and 92\,ms, which are significantly faster than most diffusion-based baselines
and are well below 0.1\,s. In contrast, ESTformer attains the smallest latency thanks to
its single-pass transformer structure, but, as shown in the main paper, does not match
SRGDiff in reconstruction quality. Overall, these results indicate that SRGDiff strikes
a favorable balance between accuracy and efficiency, and is suitable for real-time
EEG spatial super-resolution.

\begin{table}[hb]
\centering
{%
\small
\begin{tabular}{lccc}
\toprule
\textbf{Method} & \textbf{SEED} & \textbf{SEED-IV} & \textbf{Localize-MI} \\
\midrule
SasDim    & 265.4 & 269.5 & 374.4 \\
SADI      & 329.0 & 324.1 & 428.7 \\
RDPI      & 318.1 & 315.9 & 421.2 \\
DDPMEEG   & 549.3 & 558.2 & 841.7 \\
ESTformer & 3.51  & 3.55  & 5.03  \\
STAD      & 232.6 & 227.8 & 385.9 \\
SRGDiff   & 63.0  & 62.1  & 92.5  \\
\bottomrule
\end{tabular}
\caption{Runtime (ms) of different methods on SEED, SEED-IV, and Localize-MI datasets.
Values are average per-sample latency for a 4\,s EEG window measured on the same GPU.
SRGDiff remains below 0.1\,s in all cases while achieving the best reconstruction quality.}
\label{tab:runtime}
}%

\end{table}

\section{Generalization Analysis}

\subsection{Extension to Cross-Subject and Cross-Session Settings}

EEG signals are known to exhibit strong subject- and session-specific variability. To explicitly examine whether SRGDiff can generalize under this variability, we perform additional experiments on the SEED dataset in a stricter cross-subject and cross-session regime.

For the cross-session setting, we use all subjects in SEED and conduct three experiments: in each experiment, two sessions are used for training and the remaining session is held out for testing. We then report averages over all subjects and session splits. For the cross-subject setting, we train SRGDiff on subjects 1-12 and evaluate on held-out subjects 13-15 without any subject-specific fine-tuning.

The reconstruction performance under both cross-session and cross-subject protocols is summarized in Table~\ref{tab:seed_cross_generalization}. We observe that SRGDiff degrades gracefully in the cross-session setting, maintaining strong performance at $2\times$ and $4\times$ SR with more noticeable degradation at $8\times$, while the cross-subject setting is substantially more challenging and leads to larger performance drops across all SR factors than the random division settings. Nevertheless, SRGDiff maintains a clear margin over strong baselines ESTformer, STAD in terms of NMSE, PCC, and SNR, indicating that the proposed partial-observation diffusion formulation is reasonably robust to session- and subject-level variability on SEED.

\begin{table*}[!t]
\centering
\small
{%
\resizebox{\linewidth}{!}{%
\begin{tabular}{llcccccc}
\toprule
\textbf{Model} & \textbf{Metric} 
  & \multicolumn{3}{c}{\textbf{Cross-subject (SEED)}} 
  & \multicolumn{3}{c}{\textbf{Cross-session (SEED)}} \\
\cmidrule(lr){3-5} \cmidrule(lr){6-8}
& 
& \textbf{2$\times$} & \textbf{4$\times$} & \textbf{8$\times$}
& \textbf{2$\times$} & \textbf{4$\times$} & \textbf{8$\times$} \\
\midrule

\multirow{3}{*}{ESTformer}
& NMSE & 0.4411$\pm$0.004 & 0.4729$\pm$0.006 & 0.5633$\pm$0.008
        & 0.3624$\pm$0.004 & 0.4029$\pm$0.008 & 0.5129$\pm$0.007 \\
& PCC  & 0.7393$\pm$0.009 & 0.7189$\pm$0.007 & 0.6515$\pm$0.009
        & 0.7924$\pm$0.009 & 0.7742$\pm$0.007 & 0.6954$\pm$0.009 \\
& SNR  & 3.9516$\pm$0.031 & 3.5414$\pm$0.004 & 2.7279$\pm$0.039
        & 4.8753$\pm$0.034 & 4.4097$\pm$0.047 & 3.1375$\pm$0.043 \\
\midrule

\multirow{3}{*}{STAD}
& NMSE & 0.5537$\pm$0.004 & 0.7325$\pm$0.002 & 0.9267$\pm$0.008
        & 0.5617$\pm$0.003 & 0.7675$\pm$0.002 & 0.9376$\pm$0.007 \\
& PCC  & 0.6224$\pm$0.003 & 0.4587$\pm$0.002 & 0.2959$\pm$0.004
        & 0.6373$\pm$0.003 & 0.4397$\pm$0.002 & 0.2791$\pm$0.004 \\
& SNR  & 3.3872$\pm$0.097 & 1.2449$\pm$0.049 & 0.7276$\pm$0.139
        & 3.1794$\pm$0.089 & 1.1297$\pm$0.042 & 0.7168$\pm$0.122 \\
\midrule

\multirow{3}{*}{SRGDiff}
& NMSE & \textbf{0.2675}$\pm$0.003 & \textbf{0.3829}$\pm$0.005 & \textbf{0.4512}$\pm$0.005
        & \textbf{0.2480}$\pm$0.003 & \textbf{0.3529}$\pm$0.004 & \textbf{0.4127}$\pm$0.005 \\
& PCC  & \textbf{0.8023}$\pm$0.004 & \textbf{0.7232}$\pm$0.004 & \textbf{0.6902}$\pm$0.004
        & \textbf{0.8508}$\pm$0.004 & \textbf{0.7932}$\pm$0.003 & \textbf{0.7702}$\pm$0.007 \\
& SNR  & \textbf{5.6913}$\pm$0.067 & \textbf{4.5657}$\pm$0.055 & \textbf{4.1189}$\pm$0.084
        & \textbf{6.1473}$\pm$0.093 & \textbf{4.1857}$\pm$0.052 & \textbf{3.8189}$\pm$0.034 \\
\bottomrule
\end{tabular}
}
\caption{Cross-subject and cross-session reconstruction performance on SEED under different SR factors (mean $\pm$ std over folds) in terms of NMSE, PCC, and SNR.}
\label{tab:seed_cross_generalization}
}

\end{table*}

\subsection{Extension to ECoG Channel Super-resolution}

To examine whether SRGDiff is specific to scalp EEG or can generalize to other multi-channel neurophysiological signals, we further evaluate it on an invasive electrocorticography (ECoG) dataset. We use the public ECoG benchmark AJILE12 \citep{peterson2022ajile12} and follow the setting of \citet{vetter2024generating}. Applying SRGDiff here serves two purposes: (i) it tests whether our partial-observation formulation and dynamic residual guidance are \emph{modality-agnostic} within the family of spatially organized neural recordings, and (ii) it verifies that the proposed method works under a different signal regime (invasive ECoG rather than scalp EEG) without any architecture or hyperparameter changes. As shown in Table~\ref{tab:ecog_sr_generalization}, SRGDiff consistently improves over transformer-based SR baselines on the ECoG benchmark, supporting our claim that the approach extends beyond EEG to other neurophysiological channel super-resolution tasks.

\begin{table}[h]
\centering
{%
\small
\begin{tabular}{lcccc}
\toprule
Model & Metric & 2$\times$ & 4$\times$ & 8$\times$ \\
\midrule
\multirow{3}{*}{ESTformer}
 & NMSE & 0.4573 & 0.7189 & 0.8517 \\
 & PCC  & 0.7367 & 0.5299 & 0.3845 \\
 & SNR  & 3.3991 & 1.4334 & 0.6974 \\
\midrule
\multirow{3}{*}{STAD}
 & NMSE & 0.4932 & 0.6901 & 0.7987 \\
 & PCC  & 0.6854 & 0.5118 & 0.4312 \\
 & SNR  & 3.1686 & 1.4449 & 1.1684 \\
\midrule
\multirow{3}{*}{\textbf{SRGDiff (ours)}}
 & NMSE & \textbf{0.3575} & \textbf{0.6529} & \textbf{0.7312} \\
 & PCC  & \textbf{0.8023} & \textbf{0.5332} & \textbf{0.4502} \\
 & SNR  & \textbf{4.8913} & \textbf{2.1657} & \textbf{1.9089} \\
\bottomrule
\end{tabular}
\caption{Channel super-resolution performance (NMSE, PCC, and SNR) on the ECoG dataset from
\citet{vetter2024generating}, following their windowing and data split.
SRGDiff consistently improves over transformer-based baselines across all SR factors and metrics.}
\label{tab:ecog_sr_generalization}
}%

\end{table}

\subsection{Extension to Variable and Irregular LD Electrode Layouts}

In many practical deployments, the available low-density electrode layout may differ across subjects, sessions, or hardware configurations. In addition, real-world recordings often contain missing or corrupted channels, leading to irregular montages that deviate from the nominal LD design. A natural question is whether a single model can generalize across such variable and potentially irregular LD layouts.

In SRGDiff, the LD input is treated as a set of spatially localized observations that are first mapped into a common latent representation via the pretrained VAE. Concretely, each LD electrode is embedded into a continuous scalp (or cortical) coordinate space, and its signal is projected onto a fixed latent grid on which the diffusion model operates. The guidance network then consumes these latent features rather than discrete channel indices, so the conditioning is not tied to a specific LD channel configuration. This design makes the model inherently more flexible to changes in the number and spatial arrangement of LD electrodes at inference time.

To verify this empirically, we conducted two sets of experiments on SEED. First, we trained SRGDiff with a 16-electrode LD configuration and evaluated it at test time on 8/10/12/14-electrode inputs obtained by subsampling the 16-channel montage. Second, we trained SRGDiff with a 32-electrode LD configuration and evaluated it on 8- and 16-electrode inputs, again using only subsampling at test time and no retraining. As summarized in Tables~\ref{tab:seed_ld_generalization}, SRGDiff degrades gracefully as the LD montage becomes sparser: NMSE increases moderately, while PCC and SNR remain competitive across all tested LD configurations. These results indicate that a single SRGDiff model can handle sparser or irregular LD layouts at inference time without retraining, provided that the new electrodes can be embedded into the same spatial coordinate system and projected onto the latent grid used during training.

\begin{table}[t]
\centering
\small
\begin{tabular}{l l c c c}
\toprule
\textbf{Train LD chans} & \textbf{Test LD chans} & \textbf{NMSE} & \textbf{PCC} & \textbf{SNR} \\
\midrule
16        & 8          & 0.4542 & 0.7196 & 4.0329  \\
16        & 10         & 0.4031 & 0.7650 & 4.4211  \\
16        & 12         & 0.3588 & 0.8012 & 4.8705  \\
16        & 14         & 0.3245 & 0.8268 & 5.1203  \\
\textbf{16} & \textbf{16 (base)} & \textbf{0.2977} & \textbf{0.8445} & \textbf{5.2606} \\
32        & 8          & 0.4753 & 0.6735 & 3.9518  \\
32        & 16         & 0.3585 & 0.7820 & 4.4002  \\
\textbf{32} & \textbf{32 (base)} & \textbf{0.1632} & \textbf{0.9102} & \textbf{7.8413} \\
\bottomrule
\end{tabular}
\caption{SEED dataset: robustness of SRGDiff to variable LD montages. The model is trained with either a 16- or 32-channel LD configuration and evaluated on subsampled LD inputs at test time without retraining.}
\label{tab:seed_ld_generalization}
\end{table}

\color{black}
\section{Reconstruction Visualization}

\begin{figure}[h]
  \centering
  \includegraphics[width=\columnwidth]{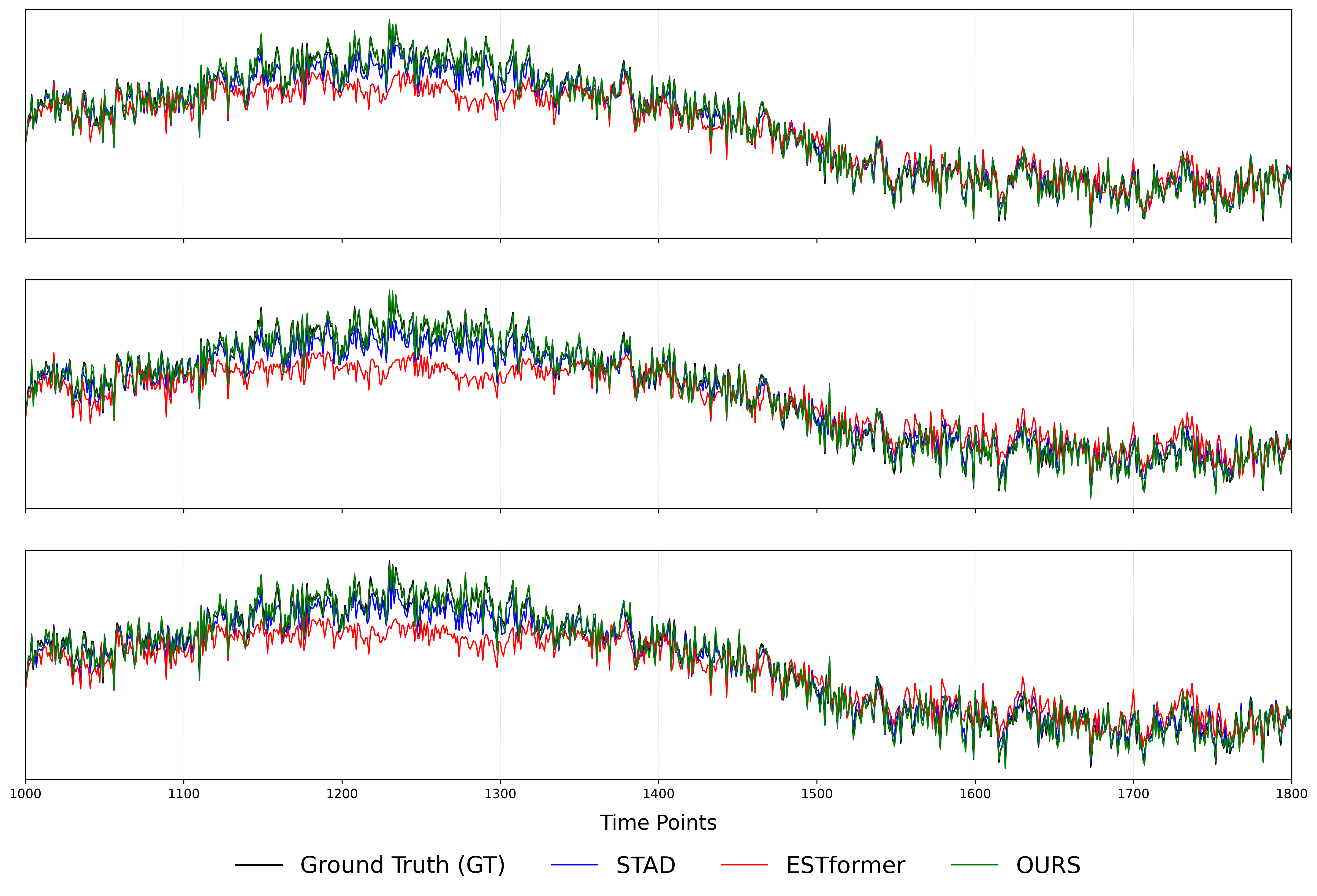}
  \caption{Reconstruction of a representative channel on Localize-MI (motor imagery): ground truth (black), STAD (blue), ESTformer (red), and SRGDiff (green).}
  \label{fig:recon_mi}
\end{figure}

\begin{figure}[h]
  \centering
  \includegraphics[width=\columnwidth]{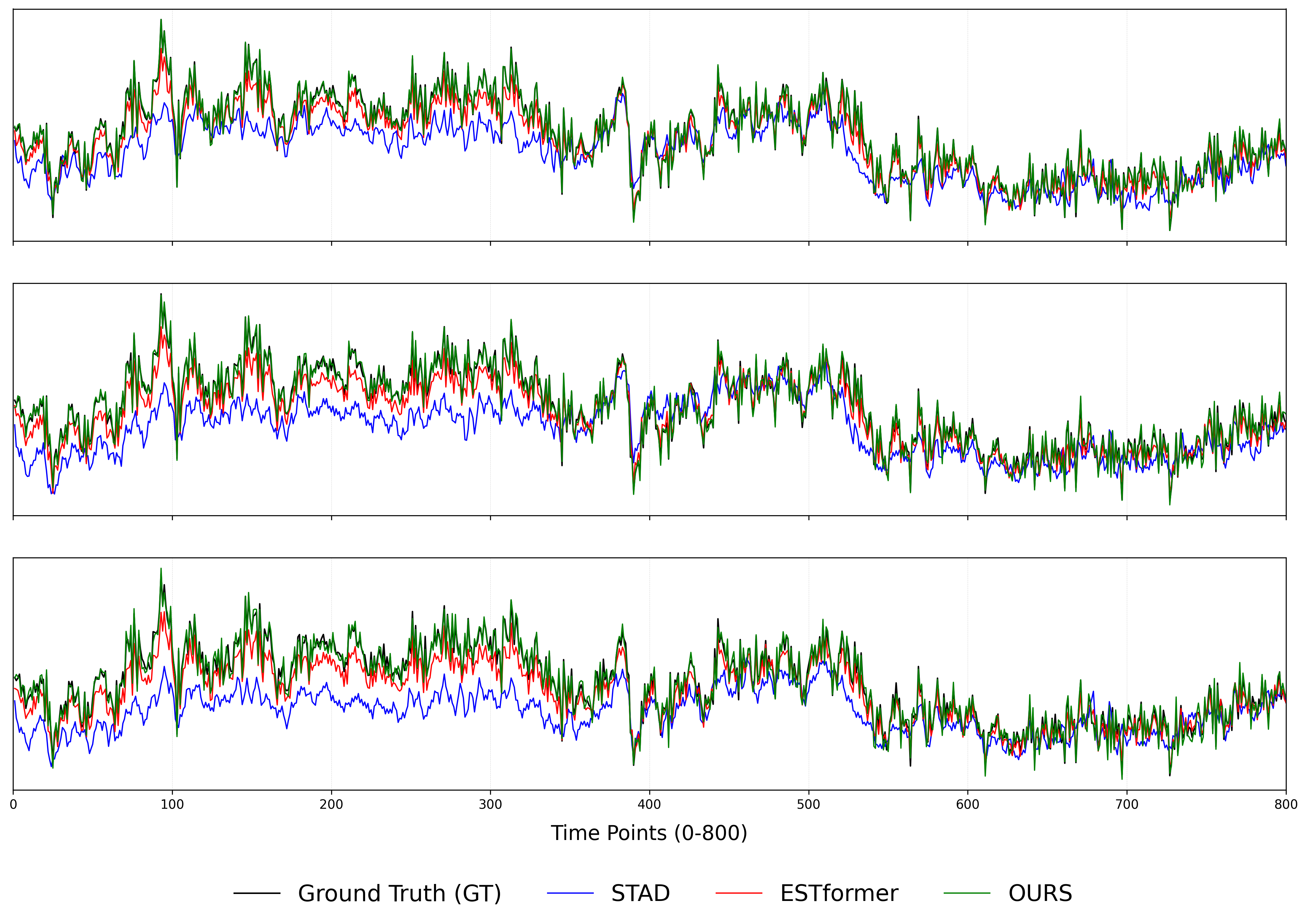}
  \caption{Reconstruction of a representative channel on SEED-IV (emotion recognition): ground truth (black), STAD (blue), ESTformer (red), and SRGDiff (green).}
  \label{fig:recon_seediv}
\end{figure}

\begin{figure}[h]
  \centering
  \includegraphics[width=\columnwidth]{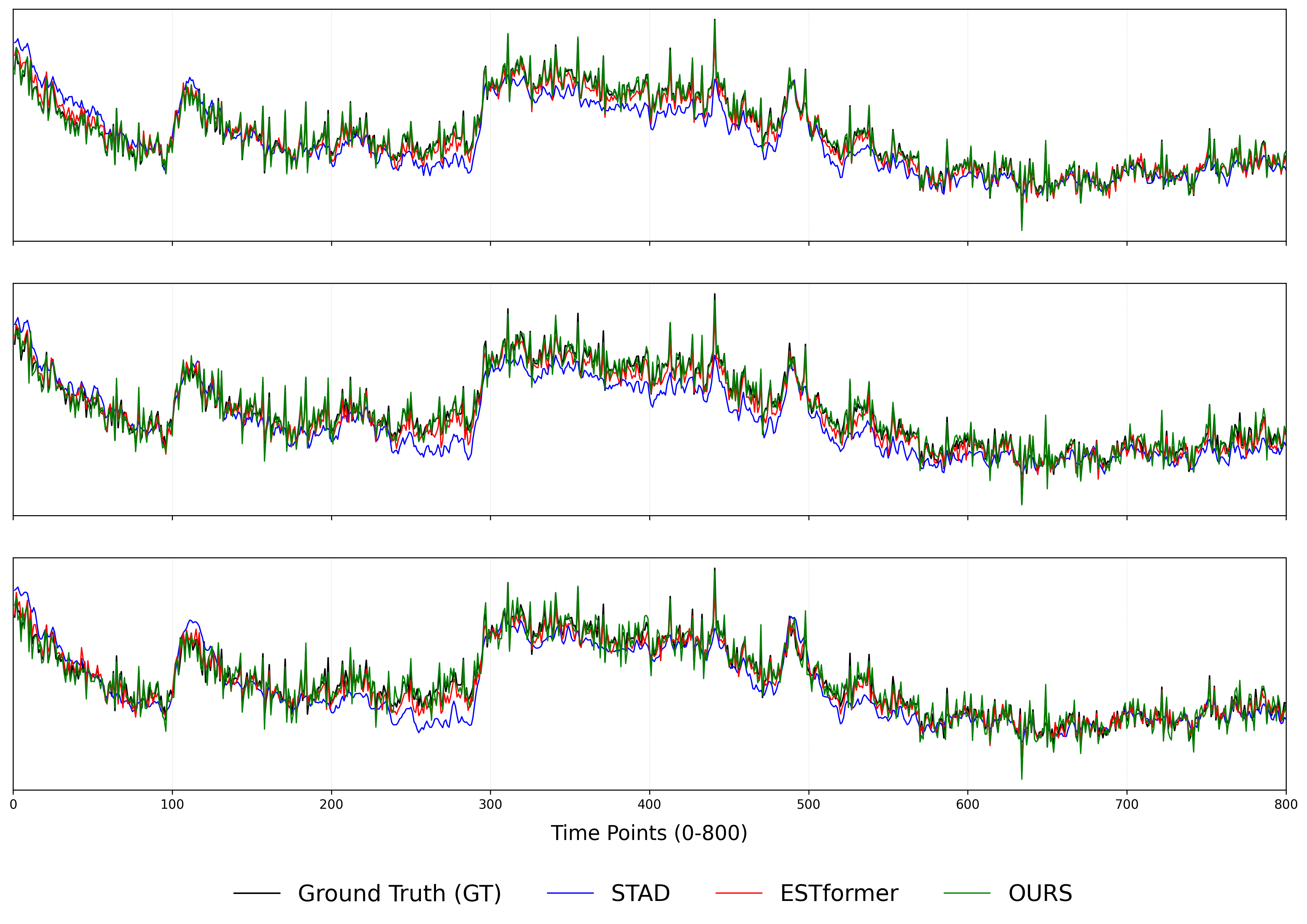}
  \caption{Reconstruction of a representative channel on SEED (emotion recognition): ground truth (black), STAD (blue), ESTformer (red), and SRGDiff (green).}
  \label{fig:recon_seed}
\end{figure}

Figure~\ref{fig:recon_mi} through Figure~\ref{fig:recon_seed} illustrate qualitative reconstructions on the three datasets. For each, we plot a single representative channel over time, comparing the ground‐truth high-density EEG (black) against STAD (blue), ESTformer (red), and SRGDiff (green). These overlays demonstrate SRGDiff’s closer alignment with the true waveform across diverse temporal patterns.

\begin{figure}[h]
  \centering
    \includegraphics[width=\columnwidth]{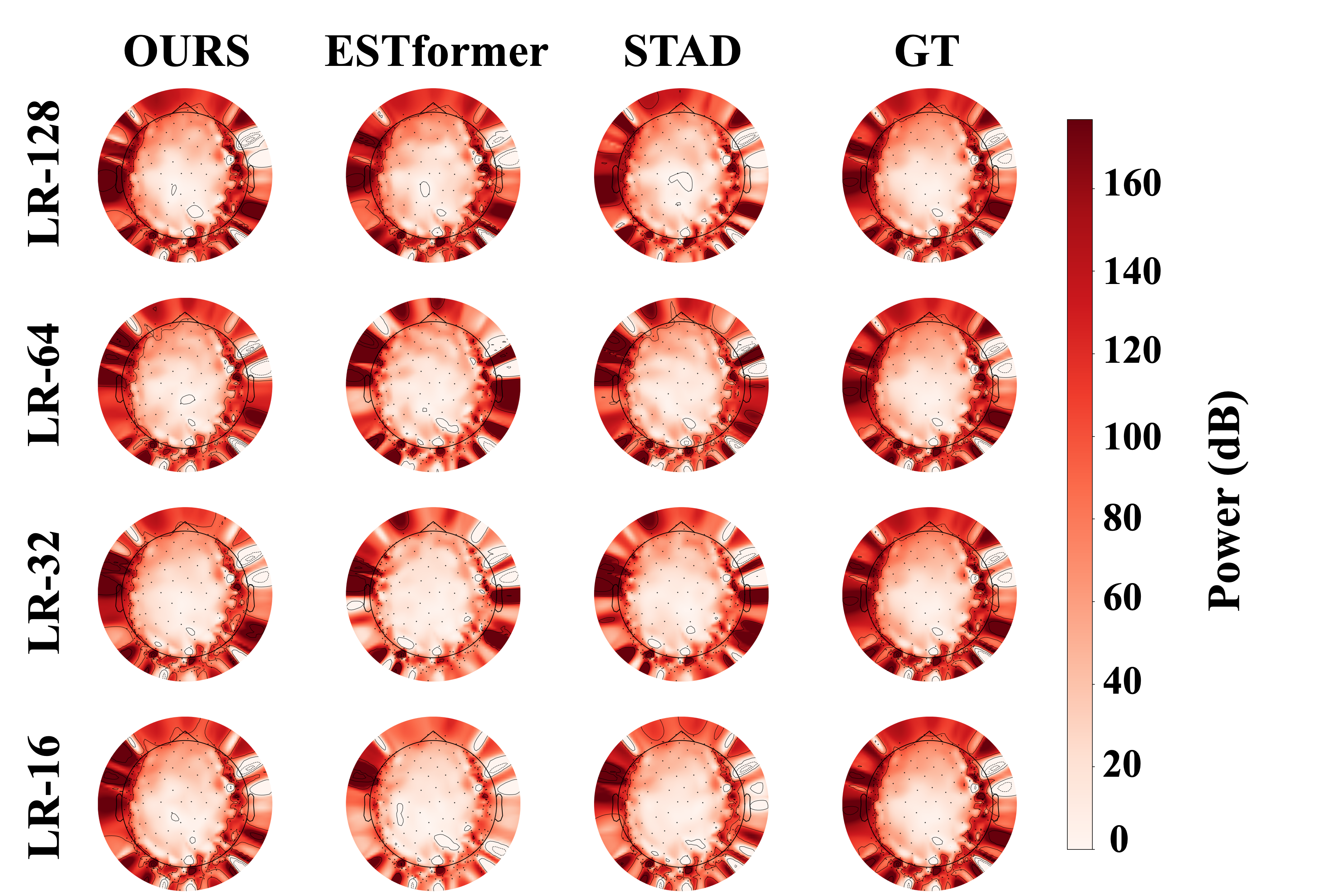}
  \caption{{Visualization of EEG topographic maps between ground-truth and reconstructed EEG signals by ESTformer, STAD and SRGDiff on Localize-MI.}}
  \label{fig:topomap_MI}
  
\end{figure}

\end{document}